\documentclass{article}
\usepackage{graphicx}
\usepackage[hypertex]{hyperref}  
\usepackage{amsmath}
\usepackage{amsfonts}
\usepackage{amssymb}
\usepackage{slashed}

\newcommand{\N}[1]{\ensuremath{\tilde{\chi}^o_#1}}

\newcommand{\GeV}{\,\,{\rm{GeV}}}
\newcommand{\fb}{\,\,{\rm{fb}}}
\newcommand{\herwig}{{\tt HERWIG}}
\newcommand{\Max}{{\tt{max}}\ }
\newcommand{\Deltam}{M_{-}}

\begin{document}

\author{Alan Barr$^{1}$ \thanks{a.barr@physics.ox.ac.uk} and Graham G. Ross$^{2}$ \thanks{g.ross@physics.ox.ac.uk}
and Mario Serna$^{2}$ \thanks{serna@physics.ox.ac.uk} \\
\small $^{1}$ Department of Physics, Denys Wilkinson Building,
University of Oxford, Keble Road, Oxford OX1 3RH,\\
\small $^{2}$ Rudolf Peierls Centre for Theoretical Physics,
University of Oxford, Keble Road, Oxford, OX1 3NP \\  \normalsize
}

\normalsize
\title{ \Large The Precision Determination of Invisible-Particle Masses at the LHC}
\normalsize

\maketitle

\begin{abstract}
We develop techniques to determine the mass scale of invisible particles pair-produced at hadron colliders.
We employ the constrained mass variable $m_{2C}$, which provides an event-by-event lower-bound to the mass scale given a mass difference.
We complement this variable with a new variable $m_{2C,UB}$ which provides an additional
{\em upper} bound to the mass scale, and demonstrate its utility with a realistic case study of a supersymmetry model.
These variables together effectively quantify the `kink' in the function $\Max m_{T2}$ which has been proposed as a mass-determination technique for collider-produced dark matter.
An important advantage of the $m_{2C}$ method is that it does not rely simply on the position at the endpoint, but it uses the additional information contained in events which lie far from the endpoint.
We found the mass by comparing the \herwig\ generated $m_{2C}$ distribution to ideal distributions for different masses.
We find that for the case studied, with $100 \fb^{-1}$ of integrated luminosity (about $400$ signal events), the invisible particle's mass can be measured to a precision of $4.1 \GeV$.
We conclude that this technique's precision and accuracy is as good as, if not better than, the best known techniques for invisible-particle mass-determination at hadron colliders.
\end{abstract}



\section{Introduction}


If dark matter is produced at a hadron collider, its likely signature will be missing transverse momentum.
To help determine the underlying origin of the observed dark-matter, it is important to measure the masses of the new particle states.
Mass determination is a key part
to identifying the underlying theory which lies beyond the Standard Model. Newly discovered particles could be Kaluza-Klein (KK) states from extra-dimensions, supersymmetric partners of known states,
technicolor hadrons, or something else that we have not anticipated.  Models predict relationships between parameters: Supersymmetry relates the couplings of current fermions
to the couplings of new bosons and the supersymmetric particle masses reflect the origin of supersymmetry breaking; masses of KK states tell us about the size of the extra dimensions.
In general, mass determination of new particle states is central to discerning what lies beyond the Standard Model.

If dark-matter particle states are discovered at the LHC, how will the masses of these new particles be measured?
Determining the mass of the dark-matter particle (or other states which decay to dark matter) will be difficult because we don't
expect the dark-matter particles to leave tracks or have a displaced vertex, and because we do not know the rest frame of the initial parton collision.

If one has confidence about the specific model and
process responsible for the observed events, then the magnitude of the process's cross-section constrains the mass.
 If one does not yet know the model, then model-independent techniques will be needed to determine properties of the newly discovered particles such as couplings, spin, phases, mixing angles, and masses.
Proposed methods for finding the gluino phase \cite{Mrenna:1999ai} or spin \cite{Kane:2008kw},
rely on our ability to make model-independent kinematic mass measurements.
 Thus, model determination needs a suite of model-independent tools to provide initial
constraints on the mass.

There has been much recent work in developing model-independent mass determination tools.
Edges in invariant mass combinations provide information about mass differences or mass squared differences  \cite{Allanach:2000kt,Allanach:2008ib}.  If the hadron collider
accesses many new states, then we may be able to determine the masses by combining the relationships provided by many different invariant-mass edges of a cascade decay
\cite{Bachacou:1999zb,Gjelsten:2004ki,Lester:2006yw,Gjelsten:2006tg}. There is also a series of approaches called Mass Shell Techniques (MST)\footnote{The title MST is suggested in Ref.~\cite{Bisset:2008hm}.} where one uses an assumption about the topology and
on-shell conditions to solve for the unknown masses. One MST variant assumes a long symmetric cascade decay chain and counts which multiplets of masses have solution to the most events \cite{Cheng:2007xv}, another assumes the masses in two events must be equal \cite{Cheng:2008mg}, and another hybrid combines a MST with the information from the many edges in cascade decays \cite{Nojiri:2007pq} \footnote{The $m_{2C}$ variable is a simple
example of such a hybrid technique.}.
There has been extensive work using the so-called `stransverse mass' variable, $m_{T2}$,
to determine the mass {\em difference} between parent particle's and a dark-matter candidate particle's mass given an assumed mass for the dark-matter candidate
(Refs.~\cite{Lester:1999tx,Barr:2003rg}
have more than 45 citations).
If the final decay to the lightest state involves a three-body decay\footnote{The presence of a $\ge 3$-body decay is a sufficient but not necessary condition. Two-body decays can also display kinks
\cite{Gripaios:2007is,Barr:2007hy} provided the decaying particles have sufficiently large transverse boosts.} as shown in Fig.~\ref{FigEventTopology},
then a `kink' in the $\Max m_{T2}$\cite{Lester:1999tx,Barr:2003rg}, will occur at a position which indicates the invisible particle mass, as described by Cho, Choi Kim, Park (CCKP) \cite{Cho:2007qv,Cho:2007dh} and corroborated in Refs.~\cite{Barr:2007hy,Nojiri:2008hy}.
The `kink' in $\Max m_{T2}$ can be quantified by the constrained mass variable $m_{2C}$ that we introduced in a previous letter \cite{Ross:2007rm}.  Even in models where new invisible particles are nearly massless (like the gravitino studied in \cite{Hamaguchi:2008hy}), one would rather not just assume the mass of the lightest supersymmetric particle (LSP), which is needed as an input to the traditional $m_{T2}$ analysis, without measuring it in some model independent way.
Which approach turns out to be best is likely to depend on what scenario nature hands us, since the various techniques involve different assumptions. Having different approaches also offers the advantage of providing a system of redundant checks.


In Ref.~\cite{Ross:2007rm}, we introduced the $m_{2C}$ kinematic variable which gives an event-by-event lower
bound on the dark-matter particle's absolute mass given the mass difference between the dark matter candidate and its parent.
In this paper, we introduce a complementary variable $m_{2C,UB}$ which gives an event-by-event \emph{upper} bound on the same absolute mass.
Our study shows that this technique rivals other invisible-particle mass determination techniques in precision and accuracy.

In this paper, we provides a demonstration of the use of the variable $m_{2C}$ and $m_{2C,UB}$ in LHC conditions.
The variables $m_{2C}(\Deltam)$ and $m_{2C,UB}(\Deltam)$ give an event-by-event lower-bound and upper-bound respectively on the mass of $Y$ assuming the topology in Fig.~\ref{FigEventTopology} and the mass difference $\Deltam=m_Y-m_N$.  To get the mass difference, we use events where $Y$ decays into $N$ and two visible states
via a three-body decay
in which we can easily determine the mass difference from the end point of the visible-states
invariant-mass distribution, $m^2_{12}$.
One might also conceive of a situation with $m_{2C}$ supplementing an alternative technique that gives a tight constraint on the mass difference but may have multiple solutions or a weaker constraint on the mass scale \cite{Gjelsten:2004ki}\cite{Serna:2008zk}.
Given this mass difference and enough statistics, $m_{2C}$'s endpoint gives the mass of $Y$.
However the main advantage of the $m_{2C}$ method is that it does not rely simply on the position at the endpoint but it uses the additional information contained in events which lie far from the endpoint.  As a result it gives a mass determination using significantly fewer events and is less sensitive to energy resolution and other errors.

To illustrate the method, in this paper we study in detail the performance of the $m_{2C}$ constrained mass variable in a specific supersymmetric model.
We study events where each of the two branches have decay chains that end with a $\N{2}$ decaying to a $\N{1}$ and a pair of opposite-sign same-flavor(OSSF) leptons.
Thus the final states of interest contain four isolated leptons
(made up of two OSSF pairs) and missing transverse momentum.
Fig~\ref{FigEventTopology} defines the four momentum of the particle states with $Y=\N{2}$, $N=\N{1}$, and the OSSF pairs forming the visible particles $1-4$.
Any decay products early in the decay chains of either branch are grouped into $k$ which we generically refer to as upstream transverse momentum (UTM).
Nonzero $k$ could be the result of initial state radiation (ISR) or decays of heavier particles further up the decay chain.
Events with four leptons and missing transverse momentum have a very small Standard-Model background.
To give a detailed illustration of the $m_{2C}$ methods, we have chosen to analyze
the benchmark point P1 from \cite{VandelliTesiPhD} which corresponds to mSUGRA with $m_o=350$ GeV, $m_{1/2}=180$ GeV,
$\tan \beta=20$, ${\rm{sign}}(\mu)=+$, $A_o=0$. Our SUSY particle spectrum was calculated with ISAJET \cite{Paige:2003mg} version~7.63.
We stress that the analysis technique employed applies generically to models involving decays to a massive particle state that leaves the detector unnoticed.

A powerful feature of the $m_{2C}$ distribution is that, with some mild assumptions, the shape away from the endpoint can be entirely determined from the
unknown mass scale and quantities that are measured.   The ideal shape fit against early data therefore provides an early mass estimate for the invisible particle.
This study is meant to be a guide on how to overcome difficulties in establishing and fitting the shape: difficulties from combinatoric issues, from differing energy resolutions for the leptons, hadrons,
and missing transverse momentum, from backgrounds, and from large upstream transverse momentum (UTM) \footnote{Our references to UTM correspond to the Significant Transverse Momentum (SPT), pair production category in \cite{Barr:2007hy} where SPT indicates that the relevant pair of parent particles can be seen as recoiling against a significant transverse momentum.}.  As we shall discuss, UTM actually provides surprising benefits.

The paper is structured as follows:  In Section \ref{SecUB}, we review $m_{2C}$ and introduce the new observation that, in addition to an event-by-event lower bound on $m_Y$,  large recoil against UTM enables one also to obtain an event-by-event \emph{upper} bound on $m_Y$.  We call this quantity $m_{2C,UB}$.   Section
\ref{SecModeling} describes the modeling and simulation employed.   Section \ref{SecShapeFactors} discusses the implications of several effects on the shape of the distribution including the $m_{12}$ (in our case $m_{ll}$) distribution, the UTM distribution, the backgrounds, combinatorics, energy resolution, and missing transverse
momentum cuts.  In Section \ref{SecPerformance}, we put these factors together and estimate the performance.  We conclude in Section \ref{SecConclusions} with a discussion about the performance in comparison to previous work.


%



\section{Upper Bounds on $m_Y$ from Recoil against Upstream  Transverse Momentum}
\label{SecUB}

We will now review the definition of $m_{2C}$ as providing an event-by-event lower bound on $m_Y$.
In generalizing this framework, we find a new result that one can
also obtain an upper bound on the mass $m_Y$ when the
two parent particles $Y$ recoil against some large upstream transverse momentum $k_T$.

\subsection{Review of the Lower Bound on $m_Y$}

\begin{figure}
\centerline{\includegraphics[width=3in]{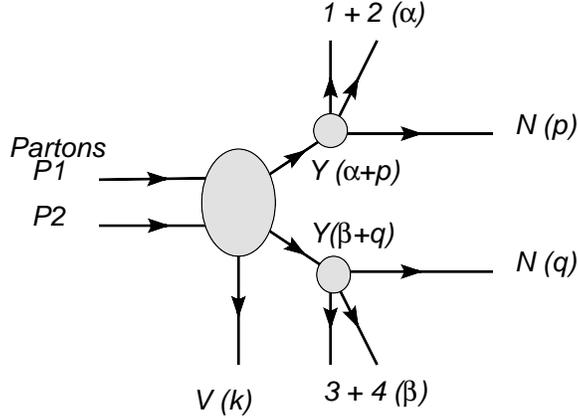}}
\caption{\label{FigEventTopology} We assume the two decay chains share a common end-state given in this diagram.  All previous decay products are grouped into the upstream transverse momentum, $k$.}
\end{figure}

Fig \ref{FigEventTopology} gives the relevant topology and the momentum assignments.  The visible particles $1$ and $2$ and invisible particle $N$ are labeled with with momentum $\alpha_1$ and $\alpha_2$ (which we group into $\alpha=\alpha_1+\alpha_2$) and $p$, respectively $\beta=\beta_1+\beta_2$ and $q$ in the other branch.  We assume that the parent particle $Y$ is the same in both branches so $(p+\alpha)^2=(q+\beta)^2$.  Any earlier decay products of either branch are grouped into the upstream transverse momentum (UTM) 4-vector momentum, $k$.

In previous work with events of this topology, we  \cite{Ross:2007rm} showed how to find an event-by-event lower bound on the true mass of $m_N$ and $m_Y$.    We assume that the mass difference $\Deltam = m_Y - m_N$ can be accurately measured from the invariant mass edges $\Max m_{12}$ or $\Max m_{34}$.
For each event, the variable $m_{2C}$ is the minimum value of the mass of $Y$ (the second lightest state) after minimizing over the unknown division of the
missing transverse momentum $\slashed{P}_T$ between the two dark-matter particles $N$:
 \begin{eqnarray}
   m^2_{2C}(\Deltam) &= & \min_{p,q}\  (p+\alpha)^2 \\
{\rm{with}\ \rm{the}\ \rm{constraints}}  & &  \nonumber \\
    (k+\alpha+\beta)_T & = & -\slashed{P}_T \label{eqC1}\\
    \sqrt{(\alpha+p)^2} -     \sqrt{(p^2)}& = &\Deltam  \label{eqC2} \\
        (\alpha+p)^2 & =&  (\beta+q)^2 \label{eqC3} \\
                (p^2) & = & (q^2). \label{eqC4}
 \end{eqnarray}
There are eight unknowns corresponding to the four momentum components of $p$ and $q$ and five equations of constraint.    Because the true $p$ and $q$ must be within the domain over which we are minimizing, the function is guaranteed to be less than or equal to $m_Y$.

One way of calculating $m_{2C}$ for an event is to use $m_{T2}(\chi_N)$ \cite{Lester:1999tx,Barr:2003rg,Ross:2007rm}, which provides a lower bound on the mass of $Y$ for an assumed mass $\chi_N$ of $N$.  The true mass of $Y$ lies along the line $\chi_Y( \chi_N)=\Deltam + \chi_N$ where we use $\chi_Y$ to denote the possible masses of $Y$ and to distinguish it from the true mass of $Y$ denoted with $m_Y$.
Thus we can see that for $\chi_N$ to be compatible with an event, we must have $m_{T2}(\chi_N) \leq \chi_Y( \chi_N) = \chi_N + \Deltam$.

For a given event, if one assumes a mass $\chi_N$ for $N$, and if the inequality $m_{T2}(\chi_N) \le \chi_N + \Deltam$ is satisfied, then there is no contradiction, and the event is compatible with this value of $\chi_N$.
If however, $m_{T2}(\chi_N) > \Deltam+\chi_N$, then we have a contradiction, and the event excludes this value $\chi_N$ as a viable mass of $N$.   Using this observation, $m_{2C}$ can be found for each event by seeking an intersection between $m_{T2}(\chi_N)$ and $\chi_N+\Deltam$ \cite{Ross:2007rm}.
The lower bound on $m_Y$ is given by $m_Y \ge \Deltam +\chi_N^o$ where $\chi_N^o$ is a zero of
  \begin{eqnarray}
    g(\chi_N)=m_{T2}(\chi_N) - \chi_N - \Deltam \label{EqgChi} \nonumber \\
    {\rm{with}}\ \ \ g'(\chi_N^o) < 0. \label{EqGprime}
     \end{eqnarray}
In the case $k=0$, the extreme events analyzed in CCKP \cite{Cho:2007qv} demonstrate that $g(\chi)$ will only have one positive zero or no positive zeros, and the slope at a zero will always be negative.
For no positive zeros, the lower bound is the trivial lower bound given by $\Deltam$.
 Note that a lower bound on the value of $m_Y$ corresponds to a lower bound on the value of $m_N$.
The Appendix in Ref.~\cite{Ross:2007rm} shows that at the
zeros of $g(\chi_N)$ which satisfy Eq(\ref{EqGprime}), the momenta satisfy Eqns(\ref{eqC1}-\ref{eqC4}).

\subsection{A New Upper Bound on $m_Y$}

If there is large upstream transverse momentum (UTM) ($k_T \gtrapprox \Deltam $)
against which the system recoils, then we find a new result.
Using the $m_{T2}$ method to calculate $m_{2C}$ gives one the immediate ability to see that $m_Y$ can also have an upper bound when requiring Eqns(\ref{eqC1}-\ref{eqC4}).
This follows because for large UTM the function $g(\chi_N)$ may have two zeros\footnote{There may be regions in parameter space where function $g(\chi)$ has more than two zeros, but we have not encountered such cases in our simulations.}
which provides both an upper and a lower bound for $m_Y$ from a single event. We have also found regions of parameter space where $g(\chi)$ has a single zero but $g'(\chi_N^o) > 0$ corresponding to an upper bound on the true mass of $m_N$ ( and $m_Y$) and
only the trivial lower bound of $m_N \ge 0$.

We can can obtain some insight into the cases in which events with large UTM provides upper bounds on the mass by studying a class of extreme event with two hard jets, $j_\alpha$ and $j_\beta$ against which $Y$ recoils ($k=j_\alpha+j_\beta$).  We will describe this extreme event and solve for the regions of parameter space for which one can analytically see the intersection points giving a lower bound and/or an upper bound.
The event is extremal in that $m_{T2}(\chi_N)$, which gives a lower bound on $m_Y$, actually gives the true value of $m_Y$ when one selects $\chi_N$ equal to the true mass $m_N$.

The ideal event we consider is where a heavier state $G$ is pair-produced on shell at threshold.  For simplicity we assume the lab-frame is the collision frame.
Assume that the $G$s, initially at rest, decay into
visible massless jets $j_\alpha$, $j_\beta$ and the two $Y$ states with the decay product momenta $\alpha+p$ and $\beta+q$.
Both jets have their momenta in the same transverse plane along the negative $\hat{x}$-axis,
and both $Y$'s momentum are directed along the $\hat{x}$-axis.  Finally, in the rest frame of the two $Y$s, both decay such that the decay products visible states have their momentum $\alpha$ and $\beta$ along the $\hat{x}$-axis and both invisible massive states $N$ have their two momenta along the negative $\hat{x}$-axis. In the lab frame, the four-vectors are given by
 \begin{eqnarray}
   j_\alpha = j_\beta & = &  \frac{m_G}{2}\left(1 - \frac{m^2_Y}{m^2_G} \right) \left\{ 1 , -1 , 0, 0 \right\} \label{eqEvent1A} \\
   \alpha = \beta & = & \frac{m_G}{2}\left(1 - \frac{m^2_N}{m^2_Y} \right) \left\{ 1 , 1 ,0 ,0  \right\} \label{eqEvent1B} \\
   p=q & = & \frac{m_G}{2} \left\{
    \left( \frac{m_N^2}{m_Y^2} + \frac{m_Y^2}{m_G^2} \right),
   \left( \frac{m_N^2}{m_Y^2} - \frac{m_Y^2}{m_G^2} \right)  ,0 ,0\right\} \label{eqEvent1C}.
 \end{eqnarray}

For the event given by Eqns(\ref{eqEvent1A}-\ref{eqEvent1C}), we can exactly calculate $m_{T2}(\chi_N)$:
 \begin{equation}
   m^2_{T2}(\chi_N) = \frac{2 \,\chi_N ^2\, m_Y^4+\left(m_N^2-m_Y^2\right)
   \left(m_N^2 m_G^2-m_Y^4\right)+\left(m_Y^2-m_N^2\right)\sqrt{4\,
   m_G^2 \,\chi_N ^2 m_Y^4+\left(m_Y^4-m_N^2
   m_G^2\right)^2} }{2\,  m_Y^4}.
 \end{equation}
This is found by calculating the transverse mass for each branch while assuming $\chi_N$ to be the mass of $N$.
The value of $p_x$ is chosen so that the transverse masses of the two branches are equal.  Substituting this value back into the transverse mass of either branch gives $m_{T2}(\chi_N)$.

\begin{figure}
\centerline{\includegraphics[width=4in]{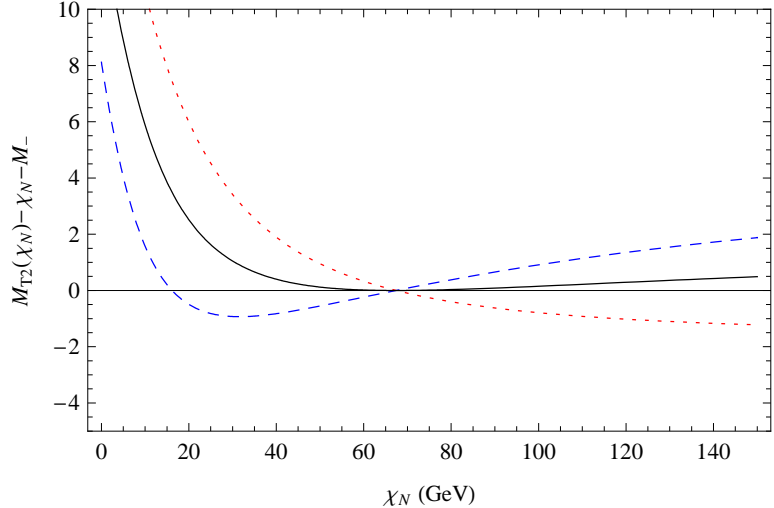}}
\caption{\label{FigExtremeMT2} Shows $g(\chi_N)$ for the extreme event in Eq(\ref{eqEvent1A}-\ref{eqEvent1C}) with $\Deltam=53 \GeV$ and $m_N=67.4 \GeV$. The dotted line has $m_G=150 \GeV$ and shows an event providing a lower bound on $m_Y$. The dashed line $m_G=170 \GeV$ and shows an event with both a lower bound and an upper-bound on $m_Y$. The the solid line shows $m_G=\sqrt{m_Y^3/m_N}$ where the lower bound equals the upper bound.}
\end{figure}

Fig \ref{FigExtremeMT2} shows $g(\chi_N)$, given in Eq(\ref{EqgChi}), for several choices of $m_G$ for the process described by Eqs(\ref{eqEvent1A}-\ref{eqEvent1C}) with $\Deltam=53 \GeV$ and $m_N=67.4 \GeV$.
Because $G$ is the parent of $Y$, we must have $m_G > m_Y$.  If $m_Y < m_G < 2 m_Y^2 /(m_N+m_Y)$, then $m_{T2}(\chi_N < m_N)$ is larger than $\chi_N + \Deltam$ up until their point of intersection at $\chi_N=m_N$. In this case their point of intersection provides a lower bound as illustrated by the dotted line in Fig.~\ref{FigExtremeMT2} for the case with $m_G=150 \GeV$.
For $ 2 m_Y^2 /(m_N+m_Y) < m_G < \sqrt{m_Y^3/m_N}$ there are two solutions
 \begin{eqnarray}
 \chi_{N,{\rm{Min}}} & = & m_N \\
 \chi_{N,{\rm{Max}}} & = & \frac{(m_N-m_Y) \left(-2 m_Y^4+m_N m_G^2 m_Y+m_N^2 m_G^2\right)}{(m_N m_G+(m_G-2 m_Y) m_Y) (m_N m_G+m_Y (m_G+2 m_Y))}
 \end{eqnarray}
When $m_G=\sqrt{m_Y^3/m_N}$, the function $g(\chi_N)$ has only one zero with the lower bound equalling the upper bound at $m_N$. The solid line in Fig.~\ref{FigExtremeMT2} shows this case.
Between $\sqrt{m_Y^3/m_N} < m_G < \sqrt{6} m_Y^2 / \sqrt{(m_N+m_Y)(2 m_N + m_Y)}$ we again have two solutions but this time with
\begin{eqnarray}
 \chi_{N,{\rm{Min}}} & = & \frac{(m_N-m_Y) \left(-2 m_Y^4+m_N m_G^2 m_Y+m_N^2 m_G^2\right)}{(m_N m_G+(m_G-2 m_Y) m_Y) (m_N m_G+m_Y (m_G+2 m_Y))} \\
 \chi_{N,{\rm{Max}}} & = & m_N.
 \end{eqnarray}
The dashed line in Fig.~\ref{FigExtremeMT2} shows this case with  $m_G=170 \GeV$.
For $m_G$ greater than this, we have $\chi_{N,{\rm{Max}}}=m_N$ and  $\chi_{N,{\rm{Min}}}=0$.

This example illustrates how $m_{2C}$ can provide both a
lower-bound and an upper-bound on the true mass
for those events with large UTM.
The upper-bound distribution provides extra information that can also be used to improve early mass determination,
and in what follows we will refer to the upper bound as $m_{2C,UB}$.  We now move on to discuss  modeling and simulation of this new observation.

\section{Modeling and Simulation}
\label{SecModeling}

As a specific example of the application of the $m_{2C}$ method, we have chosen a supersymmetry model  mSUGRA, $m_o=350$ GeV, $m_{1/2}=180$ GeV, $\tan \beta=20$, ${\rm{sign}}(\mu)=+$, $A_o=0$ \footnote{ This was model $P1$ from \cite{VandelliTesiPhD} which we also used in \cite{Ross:2007rm}.}.  The spectrum used in the simulation has $m_{\N{1}}=67.4 \GeV$ and 
$m_{\N{2}}=120.0 \GeV$.
We have employed two simulation packages.  One is a Mathematica code that creates the `ideal' distributions based only on very simple assumptions and input data.  The second is \herwig\ \cite{Corcella:2002jc,Moretti:2002eu,Marchesini:1991ch} which simulates events based on a SUSY spectrum, MSSM cross sections, decay chains, and appropriate parton distribution functions.
If the simple Mathematica simulator predicts `ideal' shapes that agree with \herwig\ generator, then one has reason to believe that
the all the relevant factors relating to the shape are identified in the simple Mathematica simulation.
This is an important check in validating the benefits of fitting the $m_{2C}$ and $m_{2C,UB}$ distribution shape as a method to measure the mass of new invisible particles produced at hadron colliders.

\subsection{Generation of ``Ideal'' Distributions}

Our `ideal' distributions are produced from a home-grown Monte Carlo event generator written in Mathematica.  This generater serves to ensure that we understand the origin of the distribution shape. It also ensures that we have control over measuring the parameters needed to determine the mass without knowing the full model, coupling coefficients, or parton distribution functions. We also use this simulation to determine on what properties the ideal distributions depends.

The simulator is used to create events satisfying the topology shown in Fig~\ref{FigEventTopology} for a set of specified masses.  We take as given the previously measured  mass difference $m_{\N{2}}-m_{\N{1}} = 52.6 \GeV$, which we use in all our simulations.
We neglect finite widths of the particle states as most are in the sub GeV range for the  model we are considering. We neglect spin correlations between the two branches.  We perform the simulations in the center-of-mass frame because $m_{2C}$ and $m_{2C,UB}$ are transverse observables and are invariant under longitudinal boosts.  The collision energy $\sqrt{s}$ is distributed according to normalized distribution
 \begin{equation}
 \rho(\sqrt{s}) = 12 \,m_{\N{2}}^2 \frac{\sqrt{s- 4\, m_{\N{2}}^2}}{s^2} \label{EqSdep}
 \end{equation}
unless otherwise specified.  The $\N{2}$ is produced with a uniform angular distribution, and all subsequent decays have uniform angular distribution in the rest frame of the parent.  The UTM is simulated by making $k_T$ equal to the UTM with $k^2=(100 \GeV)^2$ (unless otherwise specified), and boosting the other four-vectors of the event such that the total transverse momentum is zero.
As we will show, these simple assumptions capture the important elements of the process.
Being relatively model independent, they provide a means of
determining the mass for various production mechanisms.
If one were to assume detailed knowledge of the production process, it would
possible to obtain a better mass determination by using a more complete simulation
like \herwig\
to provide the `ideal' distributions against which one compares with the data.
Here we concentrate on the more model independent simulation to demonstrate that it predicts
the $m_{2C}$ and $m_{2C,UB}$ distributions well-enough to perform the mass determination that we demonstrate in this case-study.

\subsection{\herwig\  ``Data''}

\begin{figure}
\centerline{\includegraphics[width=3.1in]{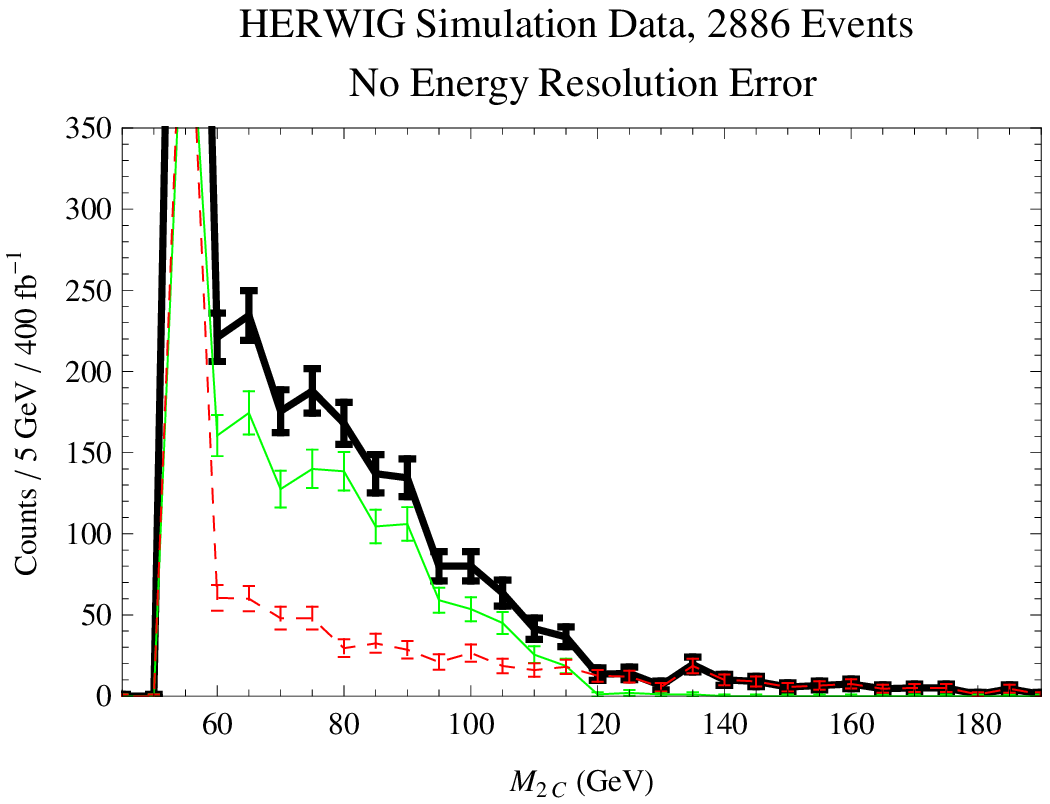}\ \includegraphics[width=3.1in]{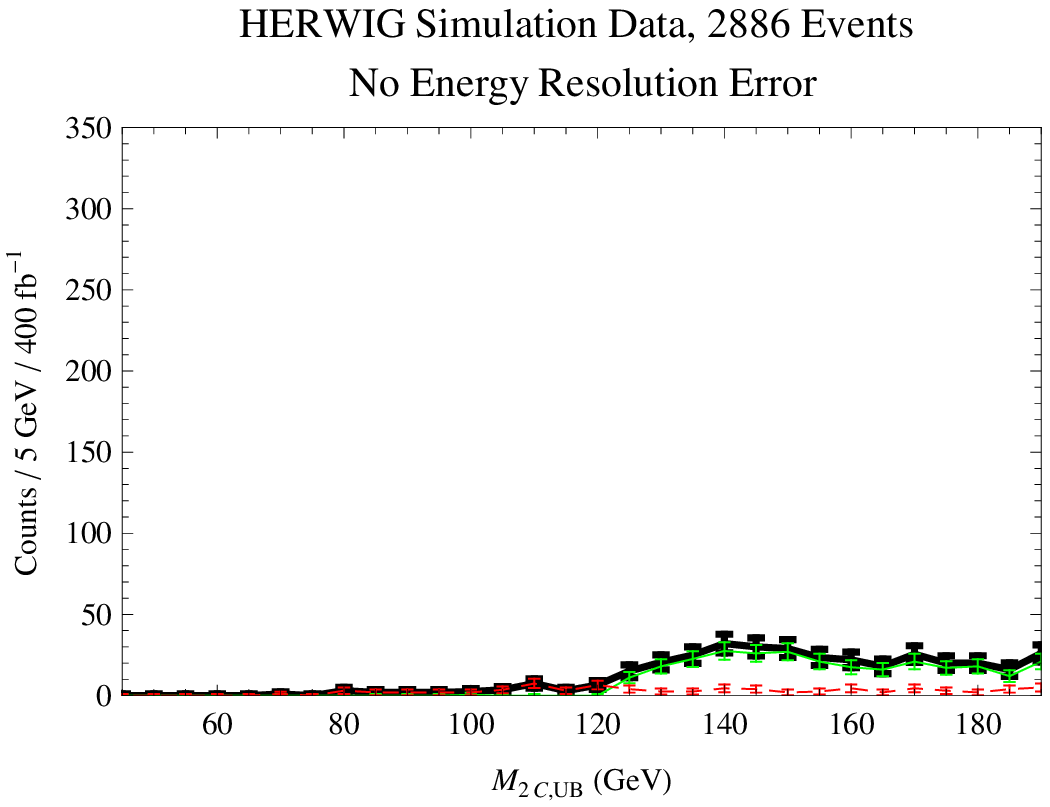}}
\caption{\label{FigHerwigResults}The $m_{2C}$ and $m_{2C,UB}$ distributions of \herwig\ events before smearing (to simulate detector resolution) is applied. The distributions' end-points show $m_{\N{2}} \approx 120$ GeV. The top thick curve shows the net distribution, the next curve down shows the contribution of only the signal events, the bottom dashed curve shows the contribution of only background events.}
\end{figure}

In order to obtain a more realistic estimate
of the problems associated with collision data,
we generate samples of unweighted inclusive supersymmetric particle pair production,
using the \herwig\ Monte Carlo program with LHC beam conditions.
These samples produce a more realistic simulation of the event
structure that would be obtained for the supersymmetric model studied here,
including the (leading order) cross sections and parton distributions.
It includes all supersymmetric processes and so contain the relevant background processes
as well as the particular decay chain that we wish to study.

Charged leptons ($e^\pm$ and $\mu^\pm$)
produced in the decay of heavy objects (SUSY particles and $W$ and $Z$ bosons)
were selected for futher study
provided they satisfied basic selection criteria on transverse momentum
($p_T>10$~GeV) and pseudorapdity ($|\eta|<2.5$).
Leptons coming from hadron decays are usually contained within hadronic jets and
so can be experimentally rejected with high efficiency using energy or track isolation criteria.
This latter category of leptons was therefore not used in this study.
The acceptance criterion used for the hadronic final state was $|\eta|<5$.
The detector energy resolution functions used are described in Section~\ref{sec:detector}.


\section{Factors for Successful Shape Fitting}
\label{SecShapeFactors}

There are several factors that control or
affect the shape of the $m_{2C}$ and $m_{2C,UB}$ distributions.  We divide the factors into those that
affect the in-principle distribution and the factors that affect
the observation of the distribution by the detector like
energy resolution and selection cuts.

The in-principle distribution of these events is influenced by
the presence or absence of spin-correlations between the branches, the
the $m_{ll}$ distribution of the visible particles, any
significant upstream transverse momentum (UTM) against  which the system is recoiling (\emph{e.g.}  gluinos or squarks decaying further up the decay chain), and background coming from other new-physics processes or the Standard Model.  As all these processes effectively occur at the interaction vertex, there are some combinatoric ambiguities.
These are the factors that influence the in-principle distribution of events that impinges on the particle detector.

The actual distribution recorded by the detector will depend on further factors.  Some factors we are able to regulate -- for example cuts on the missing transverse momentum.  Other factors depend on how well we understand the detector's operation -- such as the energy resolution and particle identification.

Where the effect of such factors is significant,
for example for the $m_{12}$, $k_T$, and background distributions,
our approach has been to model their effect on the ideal distributions
by using appropriate information from the `data', much as one would do in a real LHC experiment.
For the present our `data' are provided by \herwig, rather than LHC events, but the principle is the same.

\subsection{Factors Affecting the In-principle Distribution}

\begin{itemize}
\item \textbf{Mass Difference and Mass Scale}
\end{itemize}

\begin{figure}
\centerline{\includegraphics[width=4in]{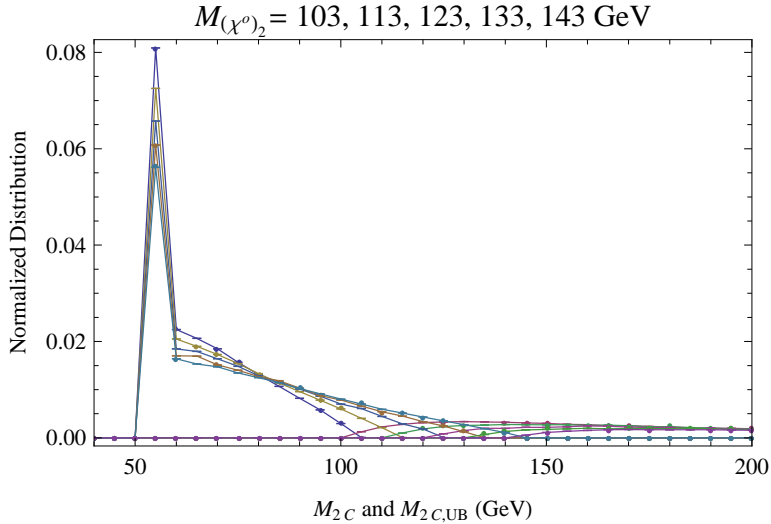}}
\caption{\label{FigLotsOfIdealCurves} We show the $m_{2C}$ and $m_{2C,UB}$ ideal
distributions for five choices of $m_{\N{2}}$
assuming the \herwig\ generated $m_{ll}$ and UTM distributions.}
\end{figure}

The end-point of $m_{2C}$ and $m_{2C,UB}$ distributions give the mass of $\N{2}$.  Therefore the mass scale, $m_{\N{2}}$, is a dominant factor
in the shape of the `ideal' distribution. This is the
reason we can use these distributions to determine the mass scale.
Fig \ref{FigLotsOfIdealCurves} shows the $m_{2C}$ and $m_{2C,UB}$ distributions
for five choices of $m_{\N{2}}$ assuming the \herwig\ generated
$m_{ll}$ and UTM distributions.

How does the shape change with mass scale?  The shape is typically sharply peaked at $m_{2C}=\Deltam$ followed by a tail that ends at the mass of $m_{\N{2}}$.
The peak at $\Deltam$ is due to events that are compatible with $m_{\N{1}}=0$.  We say these events give the trivial constraint. Because we bin the data, the height of the first bin depends on the bin size.  As $M_{+}/M_{-}=(m_{\N{2}} + m_{\N{1}})/(m_{\N{2}} - m_{\N{1}})$ becomes larger, then the non-trivial events are distributed over a wider range and the endpoint becomes less clear.  In general if all other things are equal, the larger the mass, the more events in the first bin and a longer flatter tail.

The distribution also depends on the mass difference $M_{-}$ which
we assume has been determined.
We expect that experimentally one should be able to read off the mass difference from the $m_{ll}$ kinematic end-point with very high precision.
Gjelsten, Miller, and Osland estimate this edge can be measured to better than $0.08 \GeV$  \cite{Gjelsten:2004ki,Gjelsten:2006tg}
using many different channels that
lead to the same edge, and after modeling energy resolution and background.

Errors in the mass determination
propagated from the error in the mass
difference in the limit of $k_T=0$ are given approximately by
\begin{equation}
\delta m_{\N{2}} = \frac{\delta M_{-}}{2} \left(  1- \frac{M_{+}^{2}}{M_{-}^{2}}
\right)  \ \ \ \delta m_{\N{1}} = - \frac{\delta M_{-}}{2} \left(  1+ \frac
{M_{+}^{2}}{M_{-}^{2}} \right) \label{EqDeltaMmErrorEffects}%
\end{equation}
where $\delta M_{-}$ is the error in the determination of the mass difference
$M_{-}$.  An error in  $M_{-}$ will lead to an $m_{2C}$ distribution with a shape and endpoint above or below the true mass in the direction indicated by Eq(\ref{EqDeltaMmErrorEffects}).

To isolate this source of error from the uncertainty in the fit, we assume that the mass difference is known exactly in our stated results. In our case an uncertainty of $\delta M_{-} = 0.08 \GeV$ would lead to an additional $\delta m_{\N{1}} = \pm 0.5 \GeV$ to be added in quadrature to the error from fitting.

\begin{itemize}
\item \textbf{Spin Correlations}
\end{itemize}

There are no spin correlation effects relevant to $m_{2C}$ if the $\N{2}$ pairs are directly produced and the $\N{2}$ three-body decay is dominated by the $Z$ boson (\emph{e.g} the sleptons are much heavier that $\N{2}$) \cite{Ross:2007rm}.
There are also no spin
correlations if the $\N{2}$ parents are part of a longer decay chain which involves a scalar at some stage. In the model
we are studying, there are enough vertices between the two $\N{2}$ decays that any correlation is very likely washed out, and we can treat their decays as uncorrelated.
In the simple Mathematica simulations, we have assumed no spin dependence in the production of the hypothetical ideal distribution.

\begin{itemize}
\item \textbf{Input $m_{12}$ Distributions}
\end{itemize}

\begin{figure}
\centerline{\includegraphics[width=3in]{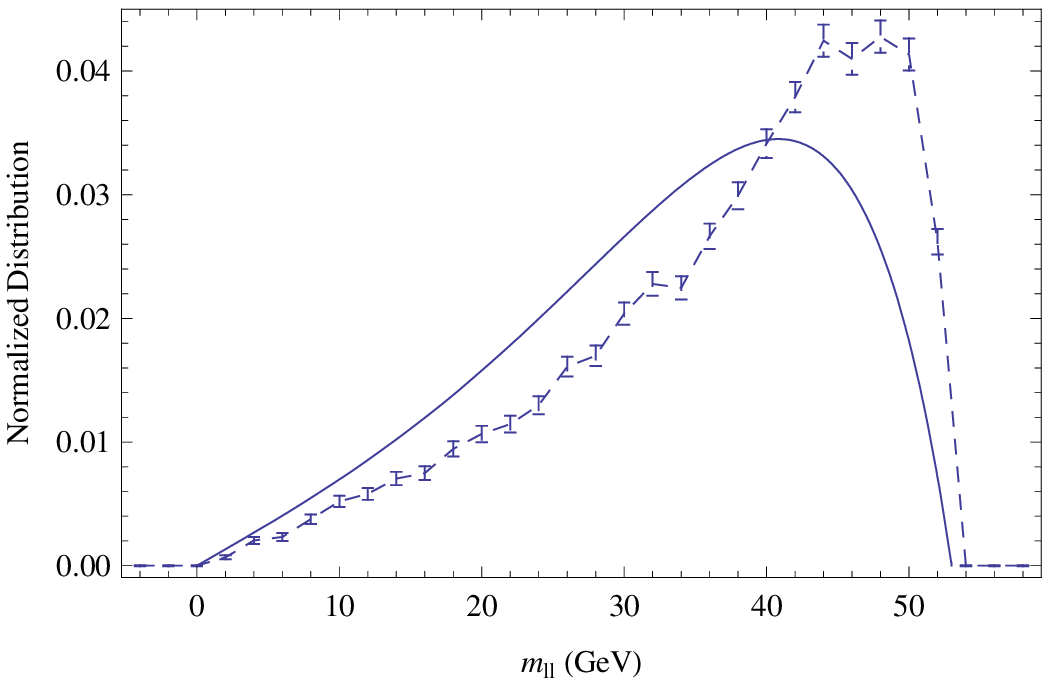} \includegraphics[width=3in]{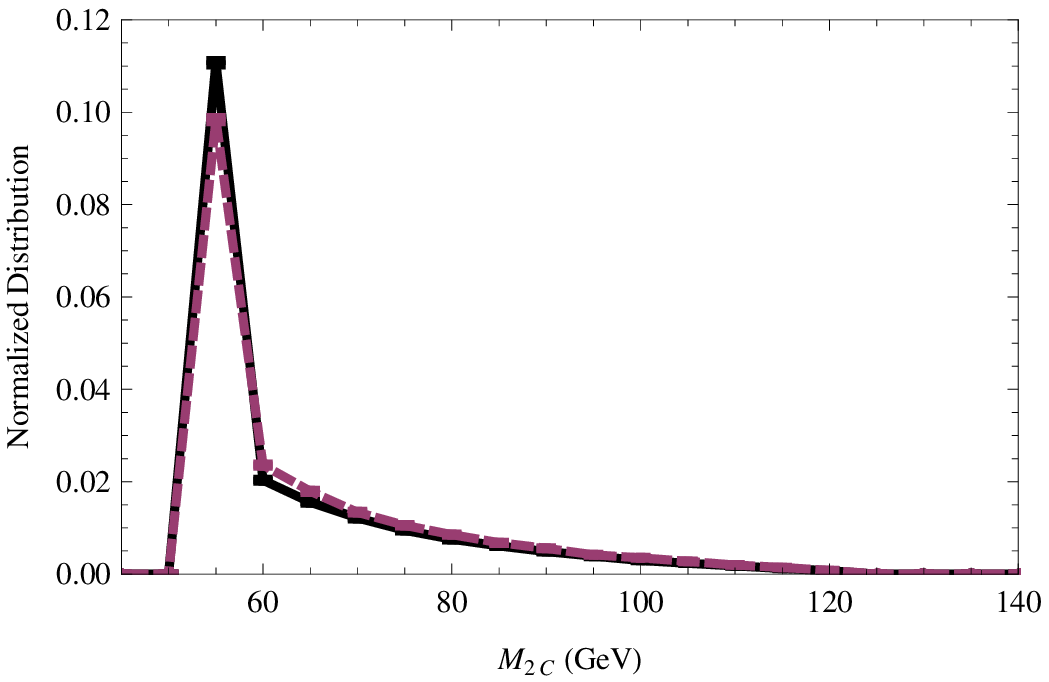}}
\caption{\label{FigMllDep} Dependence of $m_{2C}$ distribution on the $m_{ll}$ distribution with $k=0$.
{\textbf{Left:}} The $m_{ll}$ distributions. {\textbf{Right:}}  The corresponding $m_{2C}$ distributions. The solid curves show the case where the $m_{ll}$ distribution when the three-body decay is dominated by the $Z$ boson channel, and the dashed curves show the case where the $m_{ll}$ distribution is taken directly from  the \herwig\ simulation.}
\end{figure}

The $m_{ll}$ distribution affects the $m_{2C}$ distribution.
Fig \ref{FigMllDep} shows two $m_{ll}$ distributions and the corresponding $m_{2C}$ distributions with $k_T=0$ (no UTM).  The solid lines show the case where the three-body decay from $\N{2}$ to $\N{1}$ is completely dominated by a $Z$ boson.  The dashed line shows the case where the $m_{ll}$ distribution is extracted from the `realistic' \herwig\ simulation.
We can see that the $m_{2C}$ distribution
is affected most strongly in the first several non-zero bins.
If we were to determine the mass only from the shape of these first several bins
using only the $Z$ contribution for the $m_{ll}$ difference,
we would estimate of the mass to be about $4$ GeV below the true mass.
This can be understood because the the shape change of the $m_{ll}$ distribution effectively took events out of the first bin and spread them over the larger bins simulating the effect of a smaller mass.

\begin{itemize}
 \item \textbf{Input Upstream Transverse Momentum Distribution}
\end{itemize}

\begin{figure}
\centerline{\includegraphics[width=3.1in]{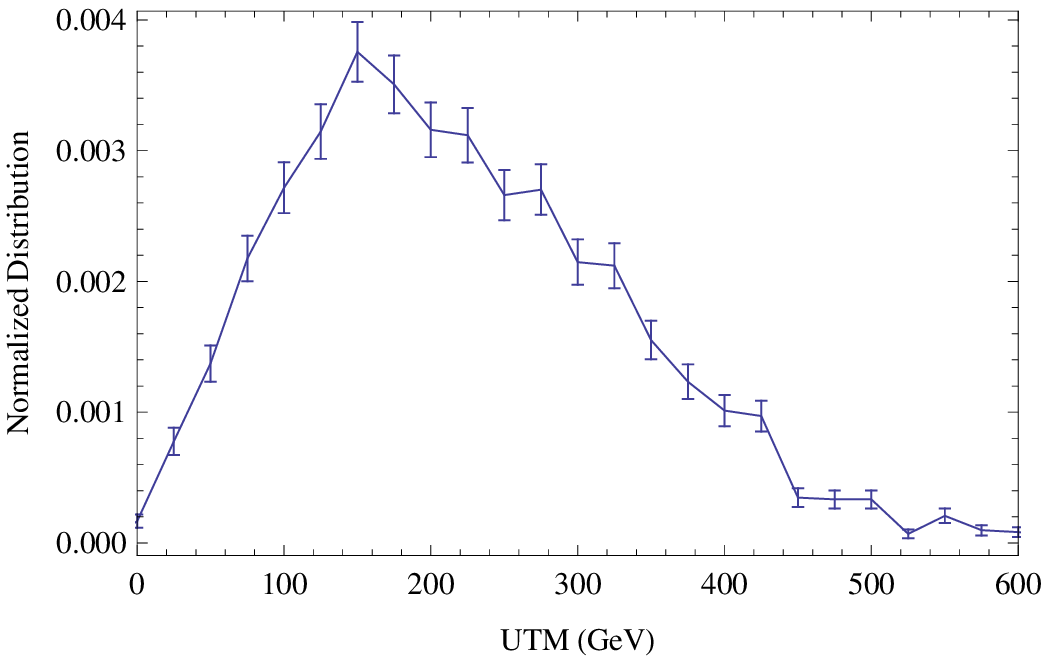}\ \includegraphics[width=3.1in]{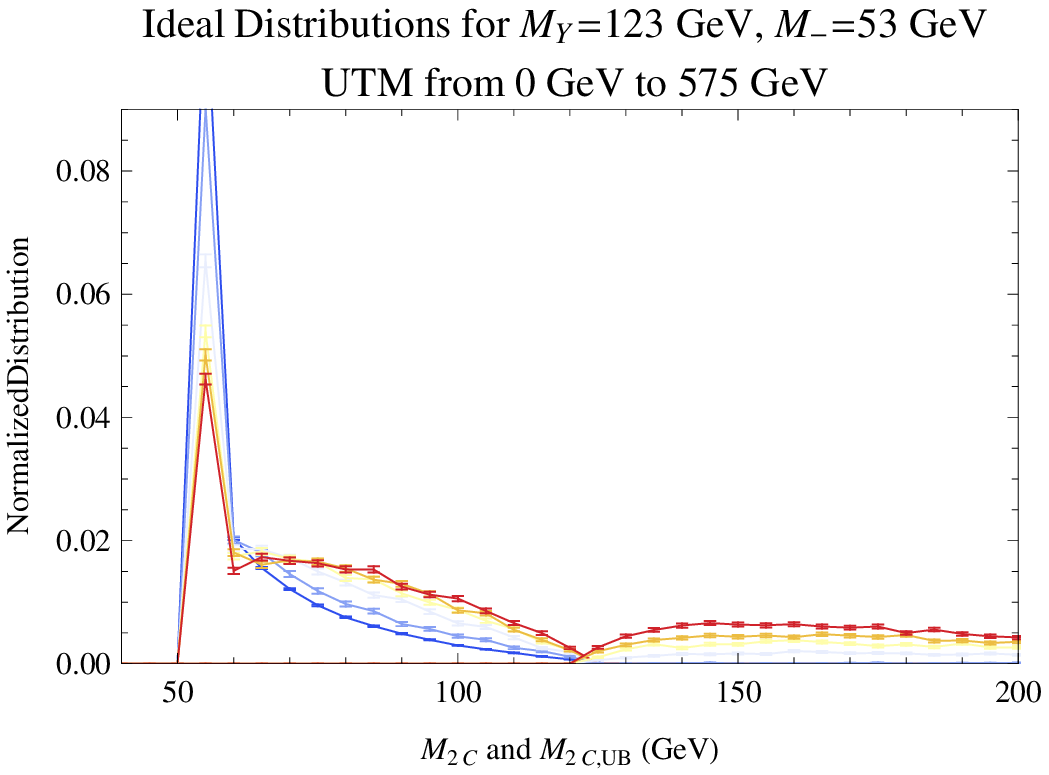} }
\caption{\label{FigM2CISRIdeal}
{\textbf{Left:}} The UTM distribution observed in the \herwig\ simulation.
{\textbf{Right:}} Ideal $m_{2C}$ upper bound and lower bound distribution for a range of upstream transverse momentum (UTM) values ($k_T = 0, 75, 175, 275, 375, 575 \GeV$) where $m_N=70$ GeV and $m_Y=123$ GeV.
}
\end{figure}

As we discussed in Section~\ref{SecUB}, if there is a large upstream transverse momentum (UTM) against which the two $\N{2}$'s recoil, then we have both an upper and lower bound on the mass scale.
The left frame of Fig.~\ref{FigM2CISRIdeal} shows the UTM distribution observed in the `realistic' \herwig\ data.
The right frame of Fig.~\ref{FigM2CISRIdeal} shows the $m_{2C}$ and $m_{2C,UB}$ distributions for fixed UTM ($k_T$) of $0$, $75$, $175$, $275$, $375$, and $575$ $\GeV$ all with $k^2=(100 \GeV)^2$. As we discuss under the next bullet, we also find the distribution is not sensitive to the value of $k^2$. For $k_T > 275 \GeV$, these curves begin to approach a common shape. These are ideal $m_{2C}$ upper and lower bound distributions where $m_{N}=70 \GeV$ and $m_{Y}=123 \GeV$.
Notice that there is no upper-bound curve for the case with zero $k_T$ UTM.
The UTM  makes the distribution have a sharper endpoint and thereby make the mass easier to determine.  This is equivalent to having a sharper kink in $\Max m_{T2}$ in the presence of large UTM \cite{Barr:2007hy}.

How do we determine $k_T$ from the data?  Because we demand exactly four leptons (two OSSF pairs), we assume all other activity, basically the hadronic activity, in the detector is UTM. 
The shape used in the `ideal' distribution is a superposition of the different fixed UTM distributions, shown on right frame of Fig.~\ref{FigM2CISRIdeal}, weighted by the observed UTM distribution, shown on the left frame of Fig.~\ref{FigM2CISRIdeal}.  Equivalently, we obtain the ideal distribution by selecting $k_T$ in the Mathematica Monte Carlo according to the observed UTM distribution.

\begin{itemize}
\item \textbf{Shape Largely Independent of Parton Distributions and Collision Energy}
\end{itemize}

In the limit where there is no UTM, then $m_{2C}$ is invariant under back-to-back boosts of the parent particles; therefore, $m_{2C}$ is also invariant to changes in the parton distribution functions.

\begin{figure}
\centerline{\includegraphics[width=3.1in]{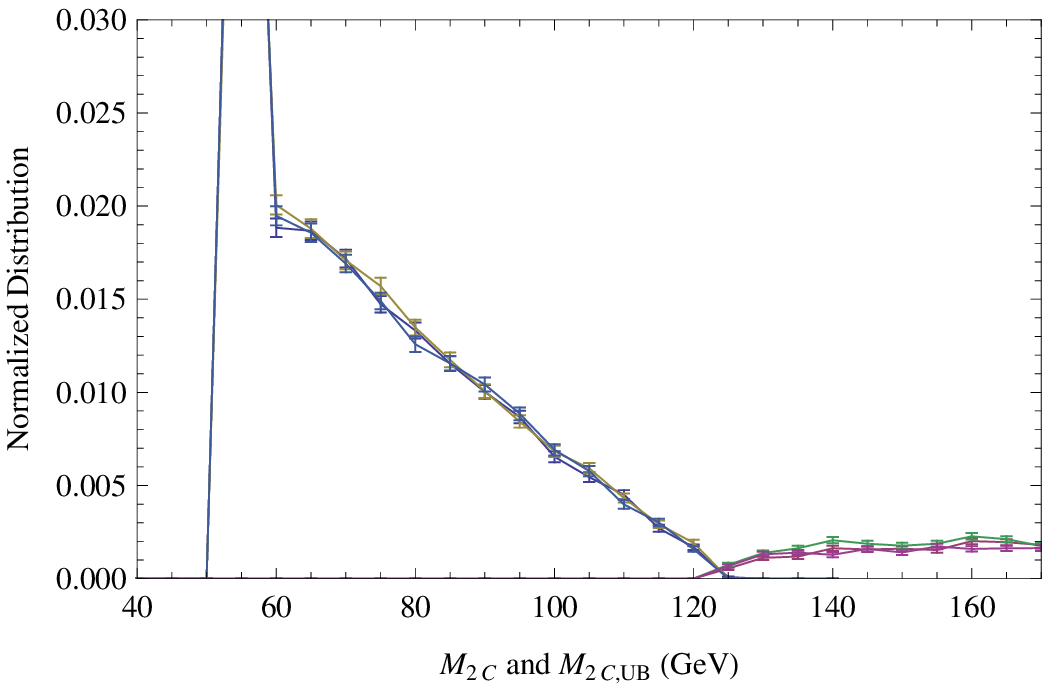}\ \includegraphics[width=3.1in]{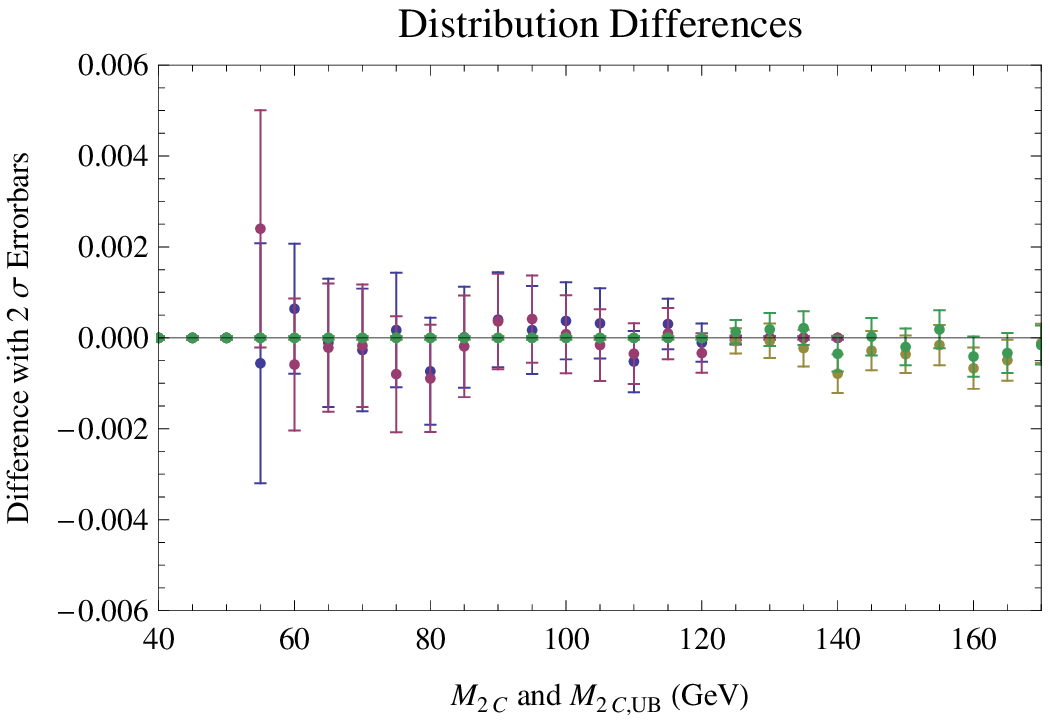}}
\caption{\label{FigDiff} Shows that even with large UTM, the distribution is independent of $k^2$ and the parton collision energy. Shown are three distributions and their difference calculated from $15000$ events. (1) $k_T=175 \GeV$, $k^2=(100 \GeV)^2$, $\sqrt{s}$ distributed via Eq(\ref{EqSdep}). (2)  $k_T=175 \GeV$, $k^2=(2000 \GeV)^2$, $\sqrt{s}$ distributed via Eq(\ref{EqSdep}). (3) $k_T=175 \GeV$, $k^2=(100 \GeV)^2$, $\sqrt{s}=549 \GeV$.}
\end{figure}

How much of this invariance survives in the presence of large UTM?  The answer is that it remains largely independent of the parton collision energy and largely independent of the mass $k^2$ as shown in Fig \ref{FigDiff} numerically.  On the left frame, we show three distributions and in the right frame their difference with $2 \ \sigma$ error bars calculated from $15000$ events. The first distribution assumes $k_T=175 \GeV$, $k^2=(100 \GeV)^2$, $\sqrt{s}$ distributed via $\ref{EqSdep}$. The second distribution assumes  $k_T=175 \GeV$, $k^2=(2000 \GeV)^2$, $\sqrt{s}$ distributed via Eq(\ref{EqSdep}). The third distribution assumes $k_T=175 \GeV$, $k^2=(100 \GeV)^2$, and a fixed collision energy of $\sqrt{s}=549 \GeV$.

\begin{itemize}
\item \textbf{Backgrounds}
\end{itemize}

\begin{figure}
\centerline{\includegraphics[width=5.0in]{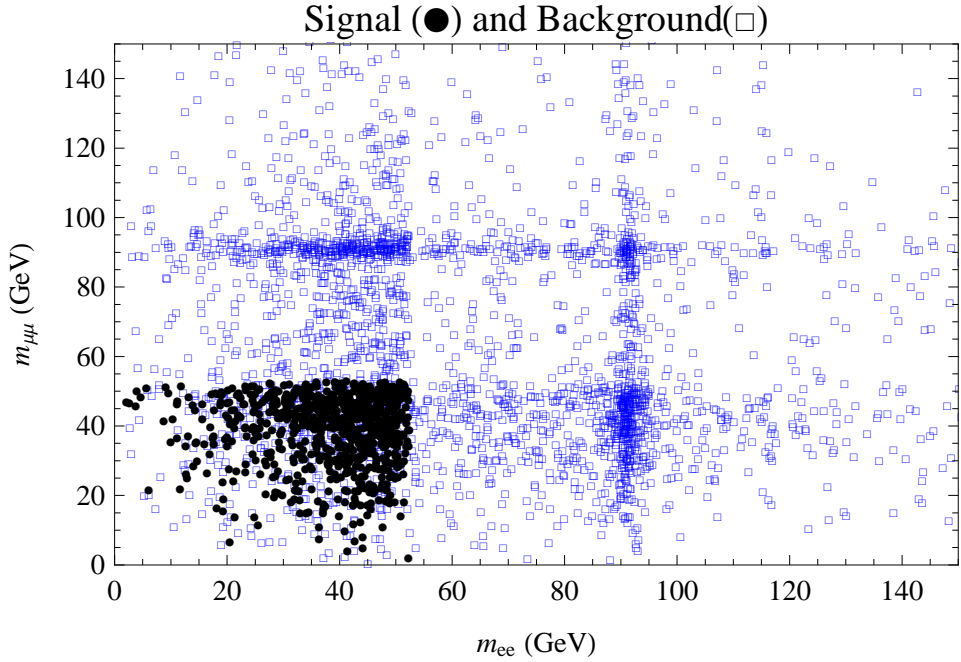}}
\caption{\label{FigMllDalitzLike} The invariant mass of the OSSF leptons from both branches of forming a Dalitz-like wedgebox analysis.
The events outside the $m_{ll} \le 53 \GeV$ signal rectangle provide
control samples from which we estimate the background shape and magnitude. The dark events are signal, the lighter events are background.}
\end{figure}

Backgrounds affect the shape and, if not corrected for,
could provide a systematic error in the estimated mass.  In Section \ref{SecPerformance} we will see that the position of the minimum $\chi^2$ in a fit to $m_{\N{1}}$ is barely affected by the background.  The main effect of the background is to shift the parabola up, giving a worse fit.  To improve the fit, we may be able to estimate the $m_{2C}$ and $m_{2C,UB}$ distribution and magnitude of the
background from the data itself.
We first discuss the sources of background, and then we describe a generic technique using a Dalitz-like wedgebox analysis to estimate a background model which gives approximately the correct shape and magnitude of the background.

One reason we study the four-lepton with missing transverse momentum channel is because of the very-low Standard-Model background \cite{Ghosh:1999ix,Bisset:2005rn}.  A previous study \cite{Bisset:2005rn} estimates about $120$ Standard-Model four-lepton events (two OSSF pairs) for $100\,\fb^{-1}$ with a $\slashed{P}_T > 20$ GeV cut. They suggest that we can further reduce the Standard-Model background by requiring several hadronic jets.  Because we expect very-little direct $\N{2}$ pair production, this would have very little effect on the number of signal events.   Also, because $Z$s are a part of the intermediate states of these background processes, very few of these events will have $m_{ll}$ significantly different from $m_Z$.

What is the source of these Standard-Model backgrounds? About $60 \%$ is from $Z$-pair production events with no invisible decay products, in which the missing transverse momentum can only arise from experimental particle identification and resolution errors.  This implies that a slightly stronger $\slashed{P}_T$ cut could further eliminate this background.   Another $40\%$ are due to $t,\bar{t}, Z$ production.
Not explicitly discussed in their study but representing another possible souce of backgrounds are events containing heavy baryons which decay leptonically.
If we assume $b$-quark hadrons decay to isolated leptons with a branching ratio of $0.01$, then LHC $t,\bar{t}$ production will lead to about $10$ events passing these cuts for $100\,\fb^{-1}$ where both OSSF leptons pairs have $m_{ll} < m_{Z}$.

Tau decays also provide a background for our specific process of interest. The process $\N{2} \rightarrow \tau^+ \tau^- \N{1}$ will be misidentified as $e^{-}e^{+}$ or $\mu^{+}\mu^{-}$ about $3\%$ of the time. Because the $\tau$ decays introduce new sources of missing transverse momentum ($\nu_\tau$), these events will distort the $m_{2C}$ calculation.  This suggests that the dominant background to the $\N{2},\N{2} \rightarrow 4 l + \slashed{P}_T + $ hadrons will be from other SUSY processes.

We now create a crude background model from which we estimate the magnitude and distribution of the background using the `true' \herwig\ data as a guide.
We follow the suggestion of Ref.~\cite{Bisset:2008hm,Bisset:2005rn} and use a wedgebox analysis plotting the invariant mass $m_{ee}$ against $m_{\mu\mu}$ to supplement our knowledge of the
background events mixed in with our signal events.  This wedgebox  analysis, seen in
Fig.~\ref{FigMllDalitzLike} for our \herwig\ simulation, shows patterns that tell about other SUSY states present. The presence of the strips along $91$ GeV indicate that particle states are being created that decay to two leptons via an on-shell $Z$.  The observation that the intensity changes above and below $m_{\mu\mu}=53$ GeV shows that many of the states produced have one branch that decays via a $\N{2}$ and the other branch decays via an on-shell $Z$.  The lack of events immediately above and to the right of the $(53 \GeV, 53 \GeV)$ coordinate but below and to the left of $(91 \GeV,91 \GeV)$ coordinate suggest that symmetric process are not responsible for this background.

We also see the density of events in the block above $53 \GeV$ and to the right of $53 \GeV$ suggest a cascade decay with an endpoint near enough to $91 \GeV$ that it is not distinguishable from $m_Z$.   Following this line of thinking, we model the background with a guess of an asymmetric set of events where one branch has new states $G$, $X$ and $N$ with masses such that the $m_{ll}$ endpoint is
\begin{equation}
  \Max m^2_{ll} ({\rm{odd}\ \rm{branch}}) = \frac{(m_G^2-m_X^2)(m_X^2-m_N^2)}{m_X^2} = (85 \GeV)^2
 \end{equation}
and the other branch is our $\N{2}$ decay.  The masses one chooses to satisfy this edge did not prove important so long as the mass differences were reasonably sized; we tried several different mass triplets ending with the LSP, and all gave similar answers.

We now describe the background model used in our fits.  One branch starts with a massive state with $m_G=160 \GeV$ which decays to a lepton and a new state $m_X=120 \GeV$ which in-turn decays to to a lepton and the LSP. The second branch has our signal decay with the $\N{2}$ decaying to $\N{1}$ and two leptons via a three-body decay. We added UTM consistent with that observed in the events.

By matching the number of events seen outside the $m_{ll} < 53 \GeV$ region, we estimate the number of the events within the signal cuts that are due to backgrounds.  We estimate $0.33$ of the events with both OSSF pairs satisfying $m_{ll} < 53 \GeV$ are background events.
The model also gives a reasonable distribution for these events.  Inspecting the actual \herwig\ results showed that actual fraction of background events was $0.4$.  If we let the fraction be free and minimize the $\chi^2$ with respect to the background fraction, we found a minimum at $0.3$.

Our background model is simplistic and does not represent the actual processes, but it does a good job of accounting for the magnitude and the shape of the background mixed into our signal distribution. Most of the \herwig\ background events came from $W$ and charginos which introduce extra sources of missing transverse momentum.  Never the less, the shape fit very accurately and the performance is discussed in Section \ref{SecPerformance}. It is encouraging that our estimate of the background shape and magnitude is relatively insensitive to details of the full spectrum.  Even ignoring the background, as we will see in Section \ref{SecPerformance}, still leads to a minimum $\chi^2$ at the correct mass.

\begin{itemize}
\item \textbf{Combinatoric Ambiguities}
\end{itemize}

If we assume that the full cascade effectively occurs at the primary vertex (no displaced vertices), then the combinatoric question is a property of the ideal distribution produced in the collisions.
There are no combinatoric issues if the two opposite-sign same-flavor lepton pairs are each different flavors.  However if all four leptons are the same flavor, we have found that we can still identify unique branch assignments 90\% of the time.  The unique identification comes from the observation that both pairs must have an invariant mass $m_{ll}$ less than the value of the $ \Max m_{ll}$ edge. In $90 \%$ of the events, there is only one combination that satisfies this requirement.  This allows one to use $95 \%$ of the four lepton events without ambiguity.  The first $50 \%$ are identified from the two OSSF pairs being of different flavors and $90 \%$ of the remaining can be identified by requiring both pairs satisfy $m_{ll} < \Max \ m_{ll}$ on which branch.  The events which remain ambiguous have two possible assignments, both of which are included with a weight of $0.5$ in the distribution.

\subsection{Factors Affecting Distribution Recorded by the Detector}
\label{sec:detector}

As just described, the `ideal' in-principle distribution is created from the observed $m_{ll}$ distribution and the observed UTM distribution.  We include combinatoric effects from events with four leptons of like flavors.  Last, we can estimate the magnitude of background events and their $m_{2C}$ and $m_{2C,UB}$ shape.
We now modify the in-principle distribution to simulate the
effects of the particle detector to form our final `ideal' distribution
that includes all anticipated effects.  The two main effects on the $m_{2C}$ and $m_{2C,UB}$ distributions are the energy resolution and the $\slashed{P}_T$ cuts.

\begin{itemize}
\item \textbf{Shape Dependence on Energy Resolution}
\end{itemize}

Energy resolution causes the $m_{2C}$ and $m_{2C,UB}$ distributions to be smeared.   Here we assume the angular resolution is negligible.
For both the Mathematica Monte Carlo model and the \herwig\ events
we simulate the detector's energy resolution by scaling the four vectors for electrons, muons, and hadrons by
\begin{eqnarray}
 \frac{\delta E_e}{E_e} & = & \frac{0.1}{\sqrt{E_e}} + \frac{0.003}{E_e} + 0.007 \\
 \frac{\delta E_\mu}{E_\mu} & = & 0.03 \\
 \frac{\delta E_H}{E_H} & = & \frac{0.58}{\sqrt{E_H}} + \frac{0.018}{E_H} + 0.025
\end{eqnarray}
respectively
\cite{Akhmadalev:2001ar}\cite{AtlasTDR}.
A more detailed detector simulation is of course possible,
but since we do not know the true behavior of any LHC detector until the device begins taking data,
a more sophisticated treatment would be of limited value here.
In practice the dependence of the ideal distribution shapes
on the missing transverse momentum resolution should reflect
the actual estimated uncertainty of the missing transverse momentum of the
observed events.

Smearing of the distributions decreases the area difference between two normalized distributions, thereby decreasing the precision with which one can determine the mass from a given number of signal events.  This expanded uncertainty can be seen in Section~\ref{SecPerformance}.

The $m_{2C}$ calculations depend on the mass difference, the four-momenta of the four leptons, and the missing transverse momentum.  As the lepton energy resolution is very tight, the missing transverse momentum's energy resolution is dominated by the hadronic energy resolution.  We model the energy resolution of the UTM as a hadronic jet.  This significantly increases the uncertainty in the missing transverse momentum because hadrons have about five times the energy resolution error.

In our Mathematica model, we represent the UTM as a single four-vector $k$, but in reality it will be the sum of many four-vectors.  Because we apply the energy resolution smearing to $k$, if $k$ is small the simple Mathematica model will have a smaller missing energy resolution.
However, an events with almost $0$ UTM could have a large missing momentum energy resolution if it has a lot of hadronic jets whose transverse momentum mostly cancels.
Fig \ref{FigM2CISRIdeal} shows that most of the time we have considerable hadronic UTM, so this effect is a minor correction on our results.


\begin{itemize}
\item \textbf{Shape Dependence on Missing Transverse Momentum Cuts}
\end{itemize}

\begin{figure}
\centerline{\includegraphics[width=5.5in]{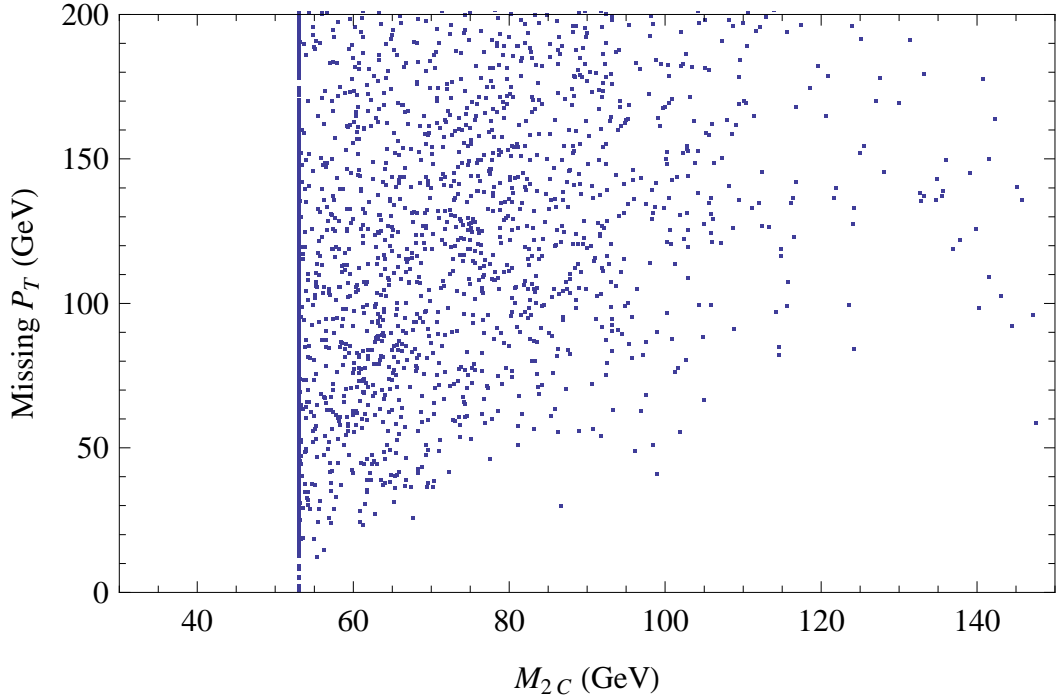}}
\caption{\label{FigPTCutDependence} The missing transverse momentum vs $m_{2C}$ values for \herwig\ data.  This shows that a $\slashed{P}_T>20 \GeV$ cut would not affect the distribution for $m_{2C} > 65 \GeV$.}
\end{figure}

A key distinguishing feature of these events is missing transverse momentum. To eliminate the large number of Standard Model events with four-lepton and with no $\slashed{P}_T$, we will need to cut on this parameter. Fig.~\ref{FigPTCutDependence} shows the \herwig\ simulation's missing transverse momentum versus the $m_{2C}$.  A non-trivial $m_{2C}$ requires substantial $\slashed{P}_T$.  Small $\slashed{P}_T$ of less than about $20 \GeV$ only affects the $m_{2C}$ shape below about $65 \GeV$.  The shape of the $m_{2C} < 65 \GeV$ therefore will require a higher fidelity model from which to train the shapes. Instead, we just choose to not fit bins with $m_{2C}< 65 \GeV$.

All events near the end of $m_{2C,UB}$ distribution require significant $\slashed{P}_T$, therefore $\slashed{P}_T$ cuts will not affect the part of this distribution which we fit.  The number of events with no non-trivial upper-bounds will also be affected by $\slashed{P}_T$ cuts.  We only fit the $m_{2C,UB}$ distribution up to about $233 \GeV$.

\section{Estimated Performance}
\label{SecPerformance}

\begin{figure}
\centerline{\includegraphics[width=3.1in]{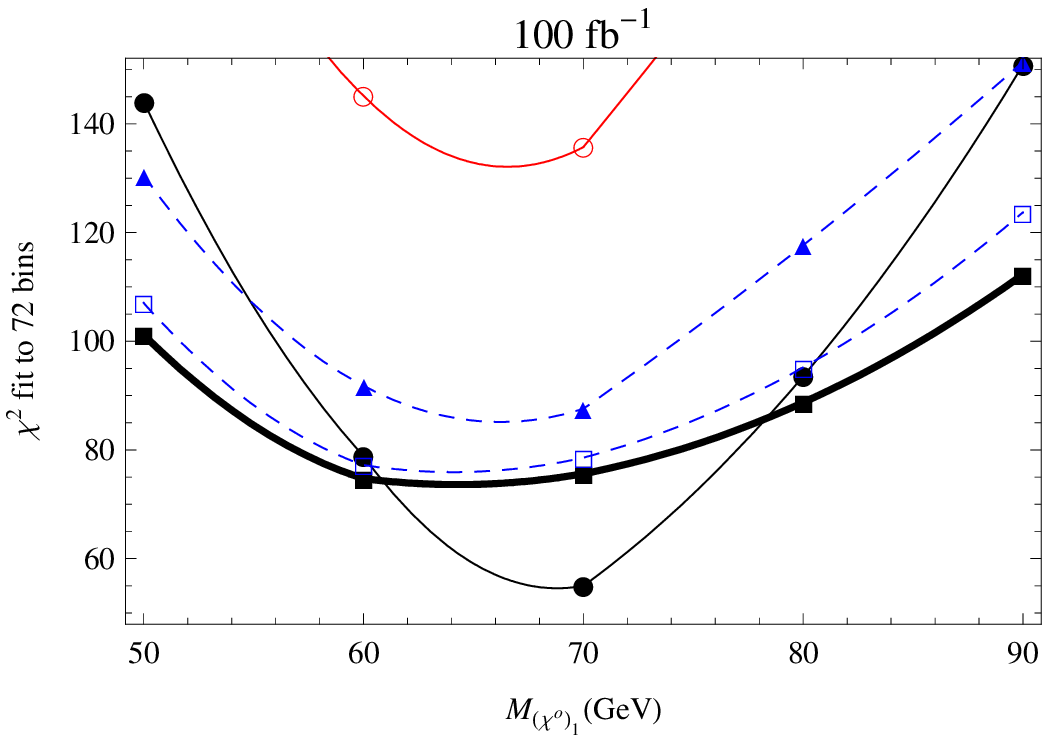}
\includegraphics[width=3.1in]{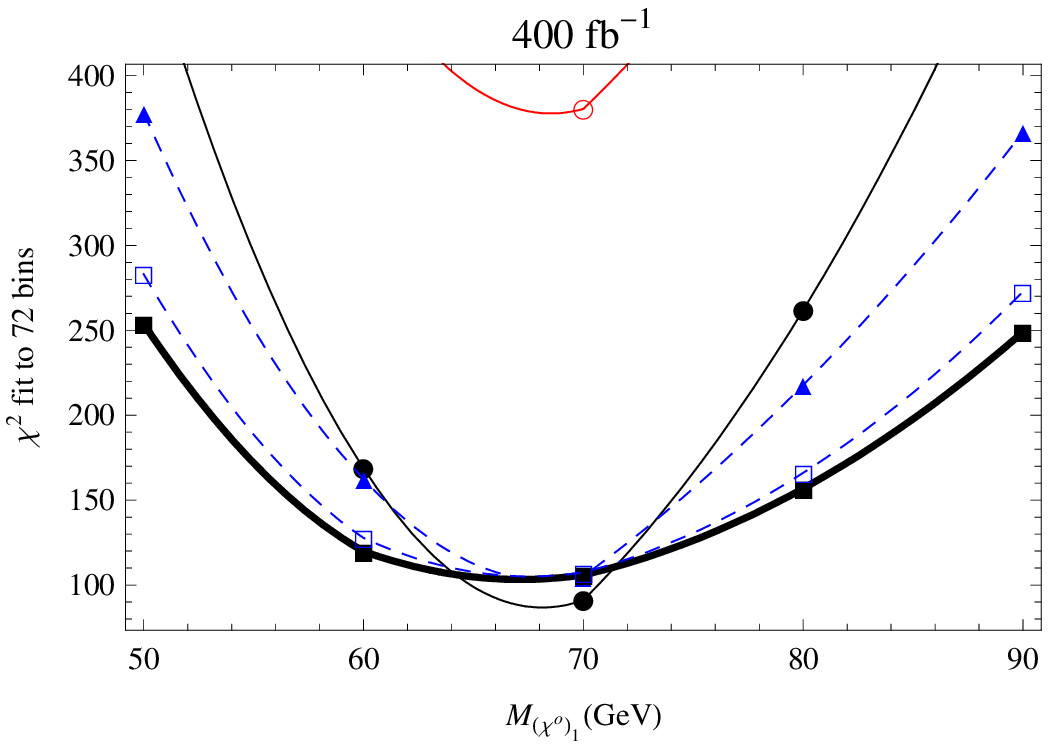}}
\caption{\label{FigPerformancePlots} The result of $\chi^2$ fits to the data with differing assumptions for $100 \fb^{-1}$ (left pane) and $400 \fb^{-1}$ (right pane). The thick line with filled squares shows the final result with all cuts, resolution error, combinatorics, and backgrounds included and estimated in the shape fitting.  This gives us $m_{\N{1}} = 63.2 \pm 4.1 \, \GeV$ with $700$ events (signal or background) representing $100 \fb^{-1}$.  After $400 \fb^{-1}$ this improves to $m_{\N{1}} = 66.0 \pm 1.8 \, \GeV$.  The error-free best case gives $m_{\N{1}} = 67.0 \pm 0.9 \, \GeV$. The correct value is $m_{\N{1}}=67.4 \GeV$.
}
\end{figure}

Determining the mass based on the shape of the distribution enables one to use all the events and not just those near the end point.
We fit both upper-bound and lower-bound shapes to the data as described in the appendix \ref{AppendixChiSqFitting}. As one expects, fitting the lower-bound shape more tightly constrains the mass from below and fitting the upper-bound shape more tightly constrains the mass from above.  Combining the two gives approximately even uncertainty.  We calculate ideal distributions assuming $m_{\N{1}}$ at five values $50$, $60$, $70$, $80$, and $90$ GeV. We then fit a quadratic interpolation through the points.  Our uncertainties are based on the value where $\chi^2$ increases by $1$ from its minimum of this interpolation.  This uncertainty estimate agrees with about $2/3$ of the results falling within that range after repeated runs.
Our uncertainty estimates do not include the error propagated from the uncertainty in the mass difference (see Eq(\ref{EqDeltaMmErrorEffects})).

We present results for an early LHC run, about $100\ {\rm{fb}}^{-1}$, and for the longest likely LHC run before an upgrade, about $400\ {\rm{fb}}^{-1}$.  After about $100\  {\rm{fb}}^{-1}$, we have $700$ events ( about $400$ signal and $300$ background).  After $400\ {\rm{fb}}^{-1}$, we have about $2700$ events (about $1600$ signal and $1100$ background).  Only $4$ events out of $1600$ are
from direct pair production.  Most of our signal events follow at the end of different decay chains starting from gluinos or squarks.  The upstream decay products produce significant UTM against which the two $\N{2}$ parent particles recoil.

First for the ideal case.  After $400\,\fb^{-1}$, using only signal events and no energy resolution, the $\chi^2$ fits to the predicted shapes give
 $m_{\N{1}} = 67.0 \pm 0.9 \, \GeV$ (filled circles in Fig.~\ref{FigPerformancePlots}).  This mass determination can practically be read off from the endpoints seen in
 Fig.~\ref{FigHerwigResults}; the $m_{2C}$ endpoint is near $120 \GeV$ and subtracting the mass differences gives $m_{\N{1}} = 120 \GeV - \Deltam =67 \GeV$. We now explore how well we can do with
 fewer events and after incorporating the effects listed in Section \ref{SecShapeFactors}.

How does background affect the fit? If we ignore the existence of background in our sample, and we fit all the events to the signal-only shapes, then we find a poor fit shown as the empty circle curve in Fig.~\ref{FigPerformancePlots}. By poor fit, we mean the $\chi^2$ is substantially larger than the $72$ bins being compared
($36$ bins from each the upper-bound and lower-bound distributions).  Despite this worse fit, the shape fits still give a very accurate mass estimate: $m_{\N{1}}=65.4 \pm 1.8 \GeV$ after $100\ {\rm{fb}}^{-1}$ and $m_{\N{1}} = 67.4 \pm 0.9 \GeV$ after $400\ {\rm{fb}}^{-1}$.
At this stage, we still assume perfect energy resolution and no missing transverse momentum cut.

Next, if we create a background model as described in Section~\ref{SecShapeFactors}, we are able to improve the $\chi^2$ fit to nearly $1$ per bin; the mass estimate remains about the same, but the uncertainty increases by about $20\%$.  We find a small systematic shift (smaller than the uncertainty) in our mass
prediction as we increase the fraction of the shape due to the background model vs the signal model.  As we increased our fraction of background, we found the mass estimate was shifted down from $66.5$ at $0\%$ background to $65.6$ when we were at $60\%$ background.  The best $\chi^2$ fit occurs with $30\%$ background; which is very close to the $33 \%$ we use from the estimate, but farther from the true background fraction of about $40 \%$. With $400 \fb^{-1}$ of data, the systematic errors are all but eliminated with the endpoint dominating the mass estimate.
These fits are shown as the triangles with dashed lines and give $m_{\N{1}} = 65.1 \pm 2.4 \GeV$ which after the full run becomes $m_{\N{1}} = 67.3 \pm 1.1 \GeV$.

Including energy resolution as described in Section \ref{SecShapeFactors} shows a large increase in the uncertainty.  The dashed-line with empty square markers shows the $\chi^2$ fit when we include both a background model and the effect of including energy resolution.
These fits are shown as the empty squares with dashed lines and give $m_{\N{1}} = 63.0 \pm 3.6 \GeV$ which after the full run becomes $m_{\N{1}} = 66.5 \pm 1.6 \GeV$.

The final shape factor that we account for are the cuts associated with the missing transverse momentum. After we apply cuts requiring $ \slashed{P}_T > 20 \GeV$ and fit only $m_{2C} > 65 \GeV$ we have our final result shown by the thick lines with filled squares. This includes all cuts, resolution error, combinatorics and backgrounds.  We find $m_{\N{1}} = 63.2 \pm 4.1 \, \GeV$ with $700$ events (signal or background)
representing $100 \fb^{-1}$, and after $400 \fb^{-1}$ this improves to $m_{\N{1}} = 66.0 \pm 1.8 \, \GeV$.
The true mass on which the \herwig\ simulation is based is $m_{\N{1}}=67.4$, so all the estimates are within about $1\ \sigma$ of the true mass.

\begin{figure}
\centerline{\includegraphics[width=4in]{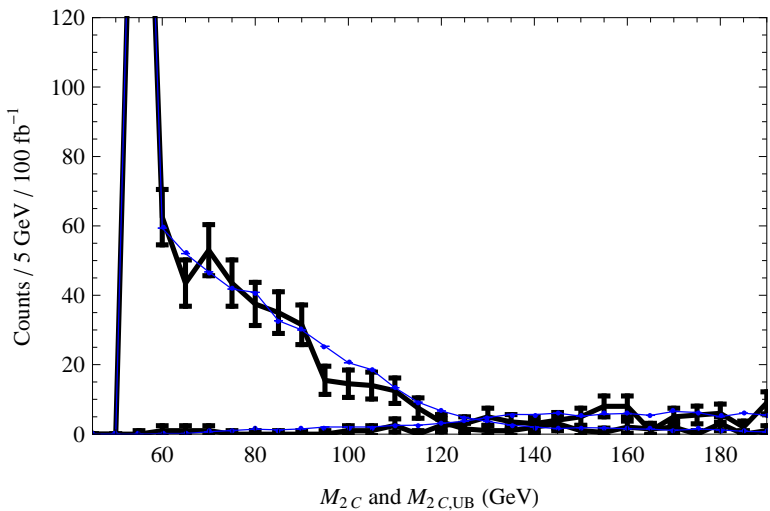}}
\caption{\label{FigHerwigWithExpectation} \herwig\ data for $100 \fb^{-1}$ (thick line) and the smooth ideal expectation assuming $m_{\N{1}}=70 \GeV$ generated by Mathematica with all resolution, background, and combinatoric effects included (thin line).
The $\chi^2$ of this curve to the \herwig\ gives the solid-black square on the left frame of Fig.~\ref{FigPerformancePlots}.}
\end{figure}
Fig \ref{FigHerwigWithExpectation} shows the ideal curve expected if $m_{\N{1}}=70 \GeV$ including all effects from energy resolution,
background, combinatoric, and $\slashed{P}_T$ cuts.
The $\chi^2$ corresponds to the solid square on the left pane of Fig.~\ref{FigPerformancePlots}.

The error in  mass determination obtained with limited statistics can be estimated using Poisson statistics.  In our studies we find that, as one would expect, increasing the number of events by a factor of four,
we bring down our error by about a factor of two.
This means that one could expect $\pm 8 \GeV$ after about $25 \fb^{-1}$ which represents $100$ signal events and $75$ background events.

\section{Discussion and Conclusions}
\label{SecConclusions}

Despite adding some of the complicating effects one would encounter with real data, we have discovered other factors which demonstrate one could obtain an even better precision than we originally reported in Ref.~\cite{Ross:2007rm}.
There, we used only our simple Mathematica model
that assumed $k=0$ and neglected most sources of realistic uncertainty.
We assumed all the events could be modeled as being direct production without spin-correlations.  With these simplifications, we argued the mass could be determined to $\pm 6 \GeV$ using $250$ signal events.


In this paper, we performed a case study to show that that the relevant  $m_{2C}$ and $m_{2C,UB}$ shapes can be successfully determined from the mass difference $\Deltam$, the $m_{ll}$ distribution observation, the upstream transverse momentum (UTM) distribution observation.
We included and accounted for many realistic effects: we modeled the large energy-resolution error of hadronic jets.  We also included the effects of backgrounds, $\slashed{P}_T$ cuts, and combinatorics.  Our signal and backgrounds were generated with \herwig.
We discussed how a Dalitz-like plot can estimate the background fraction and shape. Observed inputs were used in a simple model to determine ideal distribution shapes that makes no reference to parton distribution functions, cross sections, or other model-dependent factors that we are not likely to know early in the process.

Despite these extra sources of uncertainty, we found a final mass determination of $\pm 4.1 \GeV$ with about $400$ signal events which is still better than the appropriately scaled result from Ref.~\cite{Ross:2007rm}.
The sources of the mass determination improvement are twofold: (1) the prediction and fitting of upper-bound distribution, and (2) the sharper end-point in the presence of large UTM.
Under equivalent circumstances, the sharper endpoint is enough to give a factor of $2$ improvement in the uncertainty over the direct production case assumed in \cite{Ross:2007rm}.  Fitting the upper bound tends to improve the determination by an additional factor of $\sqrt{2}$. This improvement is then used to fight the large hadronic-jet energy resolution and background uncertainty.

Mass determination using $m_{2C}$ and $m_{2C,UB}$ applies to many other processes.  We have focused on cases where the mass difference is given by the end-point of an $m_{ll}$ distribution involving a three-body decay.
If there is not a three-body decay, then the mass difference may be found by applying other mass determination techniques like the mass shell techniques (MST) \cite{Cheng:2007xv,Cheng:2008mg,Nojiri:2007pq} or edges in cascade decays \cite{Bachacou:1999zb,Lester:2006yw,Gjelsten:2006tg} or $m_{T2}$ at different stages in symmetric decay chains \cite{Serna:2008zk}.

How does our method's performance compare to previous mass determination methods?  Firstly, this technique is more robust than the  $\Max m_{T2}$ `kink' because in fitting to the shape of the distribution, it does not rely entirely on identification of the events near kinematic boundary.  One can view $m_{2C}$ and $m_{2C,UB}$ as variables that event-by-event quantify the `kink'.  Other than the `kink' technique, the previous techniques surveyed in the introduction apply to cases where there is no three-body decay from which to measure the mass difference directly. However, each of those techniques still constrains the mass difference with great accuracy.
The technique of \cite{Bachacou:1999zb,Gjelsten:2004ki,Lester:2006yw,Gjelsten:2006tg} which uses edges from cascade decays determines the LSP mass to $\pm 3.4 \GeV$ with about 500 thousand events from $300 \fb^{-1}$.  The approach of \cite{Cheng:2008mg} assumes a pair of symmetric decay chains and assumes two events have the same structure.  They reach $\pm 2.8 \GeV$ using $700$ signal events after $300 \fb^{-1}$, but have a $2.5 \GeV$ systematic bias that needs modeling to remove.  By comparison, adjusting to $700$ signal events we achieve $\pm 2.9 \GeV$ without a systematic bias after propagating an error of $0.08 \GeV$ in the mass difference and with all discussed effects.
Uncertainty calculations differ amongst groups, some use repeated trial with new sets of Monte Carlo data, and others use $\chi^2$.
Without a direct comparison under like circumstance, the optimal method is not clear; but it is clear that fitting the $m_{2C}$ and $m_{2C,UB}$ distributions can determine the mass of
invisible particles at least as well, if not better than the other known methods in both accuracy and precision.

In summary, we have developed a mass determination technique, based on the constrained transverse mass, which is able to determine the mass of a dark-matter particle state produced at the LHC in events with large missing transverse momentum.
The $m_{2C}$ method, which bounds the mass from below, was supplemented by
a new distribution $m_{2C,UB}$ which bounds the mass from above in events with large upstream transverse momentum.
A particular advantage of the method is that it also obtains substantial
information from events away from the end point allowing for a
significant reduction in the error.
The shape of the distribution away from the end-point can be determined without detailed knowledge of the underlying model, and as such, can provide an early estimate of the mass.
Once the underlying process and model generating the event has been identified the structure away from the end-point can be improved using, for example, \herwig\ to produce the process dependent shape.
We performed a case-study simulation under LHC conditions to demonstrate that mass-determination by fitting the $m_{2C}$ and $m_{2C,UB}$ distributions survives anticipated complications.  With this fitting procedure it is possible to get an early measurement of the mass - with just 400 signal events in our case study we found we would determine $m_{\N{1}} = 63.2 \pm 4.1$.  The ultimate accuracy obtainable by this method is $m_{\N{1}} = 66.0 \pm 1.8 \GeV$.  We conclude that this technique's precision is as good as, if not better than, the best existing techniques.

\section*{Acknowledgements}

We also want to thank Chris Lester for very helpful
conversations and comments on the manuscript, and Laura Serna for reviewing the manuscript.
MS acknowledges support from
the United States Air Force Institute of Technology.
This work was partly supported by the Science and Technology Facilities Council of the United Kingdom.
The views expressed in this paper are those of the authors and do not reflect the official policy or
position of the United States Air Force, Department of Defense, or the US Government.

\appendix

\section{Appendix A: Least squares fit}
\label{AppendixChiSqFitting}

In order to determine $m_{\N{1}}$ we perform a $\chi^{2}$
fit between ideal distributions and the \herwig\ data.  First for definitions:
We define $N_{LB}$ as the number of $m_{2C}$ events in the region to be fit, and likewise
$N_{UB}$ is the number of $m_{2C,UB}$ events in the region to be fit.
The $m_{2C}$ of the events are grouped into bins; $C_{j}$ counts the events in the $j^{\rm{th}}$ bin.  The variable $f_{LB}(m_{2C\,j},m_{\N{1}})$ is the normalized $m_{2C}$ distribution of ideal events expected in bin $j$ as calculated with an assumed $m_{\N{1}}$, the measured $\Deltam$, the observed $m_{ll}$ distribution, the observed UTM distribution, and the appropriate detector simulator.  We likewise define the upper bound distribution to be $f_{UB}(m_{2C,UB\,j},m_{\N{1}})$.
We also define the background distribution for lower-bound and
upper-bound distributions to be $f_{B,LB}(m_{2C\, j})$  and $f_{B,UB}(m_{2C,UB\,j})$ and the fraction of the total events we estimate are from background $\lambda$.

Assuming a Poisson distribution, we assign an
uncertainty, $\sigma_{j}$, to each bin $j$ given by
\begin{equation}
\sigma_{LB,j}^{2}(m_{\N{1}})=\frac{1}{2}\left(  N_{LB}\,((1-\lambda)f_{LB}({m_{2C}}_{j},m_{\N{1}})+\lambda f_{B,LB}({m_{2C}}_j))  + C_{j}\right),
\end{equation}
and likewise for the upper-bound distribution.
The second term has been added to ensure an appropriate weighting of bins with very
few events that does not bias the fit towards or away from this end-point. In bins with few counts, normal Poisson statistics does not apply\footnote{By this we mean
that $N\,f({m_{2C}}_{j},m_{\N{1}})$ has a large percent error when used as a
predictor of the number of counts $C_{j}$ when $N\,f({m_{2C}}_{j},m_{\N{1}})$ is
less than about 5.}.

The $\chi^{2}$ is given by
\begin{eqnarray}
\chi^{2}(m_{\N{1}}) & = &\sum_{\mathrm{bin}\ j} \left(
\frac{C_{j}-N_{LB}\,(1-\lambda)\,f_{LB}({m_{2C}}_{j},m_{\N{1}}) - N_{LB} \,\lambda\,f_{B,LB}({m_{2C}}_j,m_{\N{1}})}{\sigma_{LB,j}}\right)  ^{2} \\
 & & + \sum_{\mathrm{bin}\ j} \left(
\frac{C_{UB,j}-N_{UB}\,(1-\lambda)\,f_{UB}({m_{2C,UB}}_{j},m_{\N{1}}) - N_{UB} \,\lambda\,f_{B,UB}({m_{2C,UB}}_j,m_{\N{1}})}{\sigma_{UB,j}}\right) ^{2}. \nonumber
\end{eqnarray}
We calculate ideal distributions for $m_{\N{1}}=50, 60, 70, 80, 90 \GeV$. We fit quadratic interplant through the points.
The minimum $\chi^{2}(m_{\N{1}})$ of the interplant is our estimate of $m_{\N{1}}$.
The amount $m_{\N{1}}$
changes for an increase in $\chi^{2}$ by one gives our $1\,\sigma$
uncertainty, $\delta m_{\N{1}}$, for $m_{\N{1}}$ \cite{Bevington}.


\begin{thebibliography}{10}

\bibitem{Mrenna:1999ai}
S.~Mrenna, G.~L. Kane, and L.-T. Wang, {\it {Measuring gaugino soft phases and
  the LSP mass at Fermilab}},  {\em Phys. Lett.} {\bf B483} (2000) 175--183,
  [\href{http://xxx.lanl.gov/abs/hep-ph/9910477}{{\tt hep-ph/9910477}}].

\bibitem{Kane:2008kw}
G.~L. Kane, A.~A. Petrov, J.~Shao, and L.-T. Wang, {\it {Initial determination
  of the spins of the gluino and squarks at LHC}},
  \href{http://xxx.lanl.gov/abs/0805.1397}{{\tt 0805.1397}}.

\bibitem{Allanach:2000kt}
B.~C. Allanach, C.~G. Lester, M.~A. Parker, and B.~R. Webber, {\it {Measuring
  sparticle masses in non-universal string inspired models at the LHC}},  {\em
  JHEP} {\bf 09} (2000) 004,
  [\href{http://xxx.lanl.gov/abs/hep-ph/0007009}{{\tt hep-ph/0007009}}].

\bibitem{Allanach:2008ib}
B.~C. Allanach, J.~P. Conlon, and C.~G. Lester, {\it {Measuring Smuon-Selectron
  Mass Splitting at the CERN LHC and Patterns of Supersymmetry Breaking}},
  {\em Phys. Rev.} {\bf D77} (2008) 076006,
  [\href{http://xxx.lanl.gov/abs/0801.3666}{{\tt 0801.3666}}].

\bibitem{Bachacou:1999zb}
H.~Bachacou, I.~Hinchliffe, and F.~E. Paige, {\it Measurements of masses in
  sugra models at lhc},  {\em Phys. Rev.} {\bf D62} (2000) 015009,
  [\href{http://xxx.lanl.gov/abs/hep-ph/9907518}{{\tt hep-ph/9907518}}].

\bibitem{Gjelsten:2004ki}
B.~K. Gjelsten, D.~J. Miller, and P.~Osland, {\it {Measurement of SUSY masses
  via cascade decays for SPS 1a}},  {\em JHEP} {\bf 12} (2004) 003,
  [\href{http://xxx.lanl.gov/abs/hep-ph/0410303}{{\tt hep-ph/0410303}}].

\bibitem{Lester:2006yw}
C.~G. Lester, {\it Constrained invariant mass distributions in cascade decays:
  The shape of the 'm(qll)-threshold' and similar distributions},  {\em Phys.
  Lett.} {\bf B655} (2007) 39--44,
  [\href{http://xxx.lanl.gov/abs/hep-ph/0603171}{{\tt hep-ph/0603171}}].

\bibitem{Gjelsten:2006tg}
B.~K. Gjelsten, D.~J. Miller, P.~Osland, and A.~R. Raklev, {\it Mass
  determination in cascade decays using shape formulas},  {\em AIP Conf. Proc.}
  {\bf 903} (2007) 257--260,
  [\href{http://xxx.lanl.gov/abs/hep-ph/0611259}{{\tt hep-ph/0611259}}].

\bibitem{Bisset:2008hm}
M.~Bisset, N.~Kersting, and R.~Lu, {\it {Improving SUSY Spectrum Determinations
  at the LHC with Wedgebox and Hidden Threshold Techniques}},
  \href{http://xxx.lanl.gov/abs/0806.2492}{{\tt 0806.2492}}.

\bibitem{Cheng:2007xv}
H.-C. Cheng, J.~F. Gunion, Z.~Han, G.~Marandella, and B.~McElrath, {\it {Mass
  Determination in SUSY-like Events with Missing Energy}},  {\em JHEP} {\bf 12}
  (2007) 076, [\href{http://xxx.lanl.gov/abs/0707.0030}{{\tt 0707.0030}}].

\bibitem{Cheng:2008mg}
H.-C. Cheng, D.~Engelhardt, J.~F. Gunion, Z.~Han, and B.~McElrath, {\it
  {Accurate Mass Determinations in Decay Chains with Missing Energy}},
  \href{http://xxx.lanl.gov/abs/0802.4290}{{\tt 0802.4290}}.

\bibitem{Nojiri:2007pq}
M.~M. Nojiri, G.~Polesello, and D.~R. Tovey, {\it {A hybrid method for
  determining SUSY particle masses at the LHC with fully identified cascade
  decays}},  {\em JHEP} {\bf 05} (2008) 014,
  [\href{http://xxx.lanl.gov/abs/0712.2718}{{\tt 0712.2718}}].

\bibitem{Lester:1999tx}
C.~G. Lester and D.~J. Summers, {\it Measuring masses of semi-invisibly
  decaying particles pair produced at hadron colliders},  {\em Phys. Lett.}
  {\bf B463} (1999) 99--103,
  [\href{http://xxx.lanl.gov/abs/hep-ph/9906349}{{\tt hep-ph/9906349}}].

\bibitem{Barr:2003rg}
A.~Barr, C.~Lester, and P.~Stephens, {\it {m(T2): The truth behind the
  glamour}},  {\em J. Phys.} {\bf G29} (2003) 2343--2363,
  [\href{http://xxx.lanl.gov/abs/hep-ph/0304226}{{\tt hep-ph/0304226}}].

\bibitem{Gripaios:2007is}
B.~Gripaios, {\it Transverse observables and mass determination at hadron
  colliders},  \href{http://xxx.lanl.gov/abs/arXiv:0709.2740 [hep-ph]}{{\tt
  arXiv:0709.2740 [hep-ph]}}.

\bibitem{Barr:2007hy}
A.~J. Barr, B.~Gripaios, and C.~G. Lester, {\it {Weighing Wimps with Kinks at
  Colliders: Invisible Particle Mass Measurements from Endpoints}},  {\em JHEP}
  {\bf 02} (2008) 014, [\href{http://xxx.lanl.gov/abs/0711.4008}{{\tt
  0711.4008}}].

\bibitem{Cho:2007qv}
W.~S. Cho, K.~Choi, Y.~G. Kim, and C.~B. Park, {\it {Gluino Stransverse Mass}},
   {\em Phys. Rev. Lett.} {\bf 100} (2008) 171801,
  [\href{http://xxx.lanl.gov/abs/0709.0288}{{\tt 0709.0288}}].

\bibitem{Cho:2007dh}
W.~S. Cho, K.~Choi, Y.~G. Kim, and C.~B. Park, {\it {Measuring superparticle
  masses at hadron collider using the transverse mass kink}},  {\em JHEP} {\bf
  02} (2008) 035, [\href{http://xxx.lanl.gov/abs/0711.4526}{{\tt 0711.4526}}].

\bibitem{Nojiri:2008hy}
M.~M. Nojiri, Y.~Shimizu, S.~Okada, and K.~Kawagoe, {\it {Inclusive transverse
  mass analysis for squark and gluino mass determination}},
  \href{http://xxx.lanl.gov/abs/0802.2412}{{\tt 0802.2412}}.

\bibitem{Ross:2007rm}
G.~G. Ross and M.~Serna, {\it {Mass Determination of New States at Hadron
  Colliders}},  {\em Physics Letters B} (2008)
  [\href{http://xxx.lanl.gov/abs/0712.0943}{{\tt 0712.0943}}]. In-press.

\bibitem{Hamaguchi:2008hy}
K.~Hamaguchi, E.~Nakamura, and S.~Shirai, {\it {A Measurement of Neutralino
  Mass at the LHC in Light Gravitino Scenarios}},
  \href{http://xxx.lanl.gov/abs/0805.2502}{{\tt 0805.2502}}.

\bibitem{Serna:2008zk}
M.~Serna, {\it {A short comparison between $m_{T2}$ and $m_{CT}$}},  {\em JHEP}
  {\bf 06} (2008) 004, [\href{http://xxx.lanl.gov/abs/0804.3344}{{\tt
  0804.3344}}].

\bibitem{VandelliTesiPhD}
W.~Vandelli, {\em Prospects for the detection of chargino-neutralino direct
  production with ATLAS detector at the LHC}.
\newblock PhD thesis.

\bibitem{Paige:2003mg}
F.~E. Paige, S.~D. Protopopescu, H.~Baer, and X.~Tata, {\it {ISAJET 7.69: A
  Monte Carlo event generator for p p, anti-p p, and e+ e- reactions}},
  \href{http://xxx.lanl.gov/abs/hep-ph/0312045}{{\tt hep-ph/0312045}}.

\bibitem{Corcella:2002jc}
G.~Corcella {\em et.~al.}, {\it {HERWIG 6.5 release note}},
  \href{http://xxx.lanl.gov/abs/hep-ph/0210213}{{\tt hep-ph/0210213}}.

\bibitem{Moretti:2002eu}
S.~Moretti, K.~Odagiri, P.~Richardson, M.~H. Seymour, and B.~R. Webber, {\it
  {Implementation of supersymmetric processes in the HERWIG event generator}},
  {\em JHEP} {\bf 04} (2002) 028,
  [\href{http://xxx.lanl.gov/abs/hep-ph/0204123}{{\tt hep-ph/0204123}}].

\bibitem{Marchesini:1991ch}
G.~Marchesini {\em et.~al.}, {\it {HERWIG: A Monte Carlo event generator for
  simulating hadron emission reactions with interfering gluons. Version 5.1 -
  April 1991}},  {\em Comput. Phys. Commun.} {\bf 67} (1992) 465--508.

\bibitem{Ghosh:1999ix}
D.~K. Ghosh, R.~M. Godbole, and S.~Raychaudhuri, {\it Signals for
  r-parity-violating supersymmetry at a 500-gev e+ e- collider},
  \href{http://xxx.lanl.gov/abs/hep-ph/9904233}{{\tt hep-ph/9904233}}.

\bibitem{Bisset:2005rn}
M.~Bisset, N.~Kersting, J.~Li, F.~Moortgat, and Q.~Xie, {\it Pair-produced
  heavy particle topologies: Mssm neutralino properties at the lhc from gluino
  / squark cascade decays},  {\em Eur. Phys. J.} {\bf C45} (2006) 477--492,
  [\href{http://xxx.lanl.gov/abs/hep-ph/0501157}{{\tt hep-ph/0501157}}].

\bibitem{Akhmadalev:2001ar}
{\bf ATLAS} Collaboration, S.~Akhmadalev {\em et.~al.}, {\it {Hadron energy
  reconstruction for the ATLAS calorimetry in the framework of the
  non-parametrical method}},  {\em Nucl. Instrum. Meth.} {\bf A480} (2002)
  508--523, [\href{http://xxx.lanl.gov/abs/hep-ex/0104002}{{\tt
  hep-ex/0104002}}].

\bibitem{AtlasTDR}
A.~Collaboration, {\it {ATLAS} computing : Technical design report},  {\em
  {CERN}, {ATLAS-TDR-017},{CERN-LHCC-2005-022}} (2005).

\bibitem{Bevington}
P.~Bevington and K.~Robinson, {\em Data Reduction and Error Analysis in the
  Physics Sciences}.
\newblock McGraw Hill, second edition~ed., 1992.

\end{thebibliography}

\providecommand{\href}[2]{#2}\begingroup\raggedright\endgroup

\end{document}